\documentclass[final,3p]{elsarticle}
\usepackage{setspace}

\usepackage{subfigure}
\usepackage{lineno,hyperref}
\usepackage{amsmath,amssymb,amsfonts}
\usepackage{algorithmic}
\usepackage{textcomp}
\usepackage{xcolor}
\usepackage{amsthm}
\usepackage{steinmetz}
\usepackage{graphicx,epstopdf}
\usepackage{breqn}
\usepackage{amstext}
\usepackage{bm}
\usepackage{multirow}
\usepackage{float}
\usepackage{mathtools}
\usepackage{enumerate}
\usepackage{enumitem}
\usepackage{caption}
\usepackage{epstopdf}

\usepackage{hyperref}

\makeatletter
\def\makeLineNumberLeft{%
  \linenumberfont\llap{\hb@xt@\linenumberwidth{\LineNumber\hss}\hskip\linenumbersep}
  \hskip\columnwidth
  \rlap{\hskip\linenumbersep\hb@xt@\linenumberwidth{\hss\LineNumber}}\hss}
\leftlinenumbers
\makeatother

\usepackage{blindtext}

\usepackage{tikz}
\usetikzlibrary{fit,shapes.misc}

\usepackage{amssymb}

\usepackage{geometry}

\journal{arXiv}

\begin{document}

\begin{frontmatter}



\title{Real-Time Simulation of Level 1, Level 2, and Level 3 Electric Vehicle Charging Systems}



\author{Li Bao, Lingling Fan$^*$, Zhixin Miao}
\address{$^{*}$Department of Electrical Engineering, University of South Florida, Tampa FL USA 33620.}
\address{Email: linglingfan@usf.edu.}

\begin{abstract}
A charging system is required to convert ac electricity from the grid to dc electricity to charge
an electric vehicle (EV) battery. According to the Society of Automatic Engineers (SAE) standard,
EV chargers can be divided into three levels based on power rating: Level 1, Level 2, and Level
3. This paper investigates the circuit topologies and control principles of EV charging systems
at each level. Three high-fidelity testbeds of EV charging systems for a 10 kWh battery are
designed and implemented in real-time digital simulator RT-Lab. The testbeds include modeling
details such as switching of semiconductors.  Twenty-five minutes real-time simulation is
conducted for each testbed. Detailed dynamic performance of the circuits and the controls at
every stage are presented to demonstrate the charging process. All three level EV charging
systems employ high-frequency transformer embedded dual active bridge (DAB) dc/dc converter to
regulate battery side dc voltage and current. Hence, average model-based linear system analysis
is given to configure the parameters of the phase shift control adopted by the DAB dc/dc
converter. In addition, power factor control (PFC) that is employed for Level 1 and Level 2
single-phase ac charging systems, three-phase voltage source converter control that is employed
for Level 3 three-phase ac charging systems, are all analyzed. The three testbeds, with their
detailed circuit parameters and control parameters presented, can be used as reference testbeds
for EV grid integration research.
\end{abstract}


\end{frontmatter}


\section{Introduction}
\label{section1} Environmental degradation, energy security and oil depletion urge  governments and
society to use clean energy. In U.S., the transportation energy consumption accounted for $29\%$ of
total energy in 2017 \cite{USenergy}. Majority of the energy consumed is fossil oil. The large
amount consumption of oil causes air pollution by carbon dioxide emission, nitrogen oxide emission
and other poisonous gases. Therefore, the traditional transportation vehicles, such as internal
combustion engine vehicles, are expected to be reduced, while EVs are expected to have increased
usage.

EVs use electrical power stored in a battery to drive motors. The battery is charged from grid
through a charging system that converts the grid ac electricity to dc electricity. The charging
system is categorized into three levels according the SAE standard \cite{Areviewonthe}. Level 1
charger utilizes  household outlets for overnight charging. The charging voltage is consistent with
house use at $120$ V level. The charging time is about 12 hours. Level 1 charging is the slowest
charging. Level 2 charger is designed for private or public facilities such as workplace or mall.
The charging voltage is $240$ V and the current is up to $60$ A. Level 3
charging is fast charging for commercial use. Level 3 chargers are located at specific locations.
Level 3 chargers use three phase voltage sources and have over 20 kW power level. The three
charging levels are summarized in Table \ref{tab1}.

\begin{table}[h]
\caption{\label{tab1}Charging levels summarization}
\begin{center}
\begin{tabular}{cccc}
   \hline      \hline
    \multirow{1}{*} {Charging}& {Supply} & {Charging} &{Rating}\\
    {Level}&  {Voltage}&  {Current}&  {Power}\\      \hline
    Level 1 & $120$ V, single-phase & up to $16$ A& up to $1.92$ kW\\
    Level 2 & $240$ V, single-phase & up to $60$ A& up to $14.4$ kW\\
    Level 3 & Not finalized, 3-phase & Not finalized& over $20$ kW\\
    \hline\hline
\end{tabular}
\end{center}
\end{table}

In order to realize high efficiency in power transfer and convert AC grid voltage to DC voltage,
a charging system may include a rectifier to rectify ac to dc, a power factor correction (PFC)
boost dc-dc circuit to achieve unity power factor, and a dual-active-bridge (DAB) dc-dc converter
to achieve flexible dc to dc voltage conversion. The controls include DC bus voltage control, PFC
control and constant current/constant voltage (CC/CV) control.

Computer simulation is the cheapest tool to conduct experiments. Considering the complicated model
and control structure, the off-line simulation software such as PSCAD and PSpice become ill suited
due to low simulation speed and limited memory. Compared to the off-line simulation, real-time
simulation is able to provide fast, reliable and accurate simulation \cite{RTSR}.

In this paper, three testbeds representing three levels of EV charging systems will be implemented
in Real-Time Laboratory (RT-Lab), a real-time simulator developed by Opal-RT technologies
\cite{1626347,RT2,RT3}. 

The paper investigates the detailed models of charging systems, including their circuit topologies
and control structures. Principles of each control, e.g., PFC, phase shift, are examined in detail.
The models are then implemented in RT-Lab. Simulation results such as the inductor current of PFC,
DC bus voltage, phase shift of DAB's ac voltage at two ends, and battery state-of-charge (SOC) are
presented and examined. The three testbeds are compared for 25 minutes' real-time simulation of 10
kWh battery charging. To the authors best knowledge, there is no existence of real-time simulation
models for EV chargers of all three level
in the literature. 

The contribution of this paper is two-fold. (1) Simulation models of EV charging systems at three
levels are developed in RT-Lab real-time simulator in discrete time domain. (2) The charging
process and the principles of control strategies are examined and further validated through
real-time simulation results.

The rest of the paper is organized as follows. Level 1 and Level 2 charging systems are presented
in Section 2. Level 3 charging system is presented in Section 3. The RT-Lab structure and
performance are illustrated in Section 4. Section 5 demonstrates the real-time simulation results
for the three testbeds. Finally Section 6 concludes this paper.

\section{Level 1 and Level 2 charging systems}
\label{section2} 
\begin{figure*}[ht!]
\centering
\includegraphics[width=5.8in]{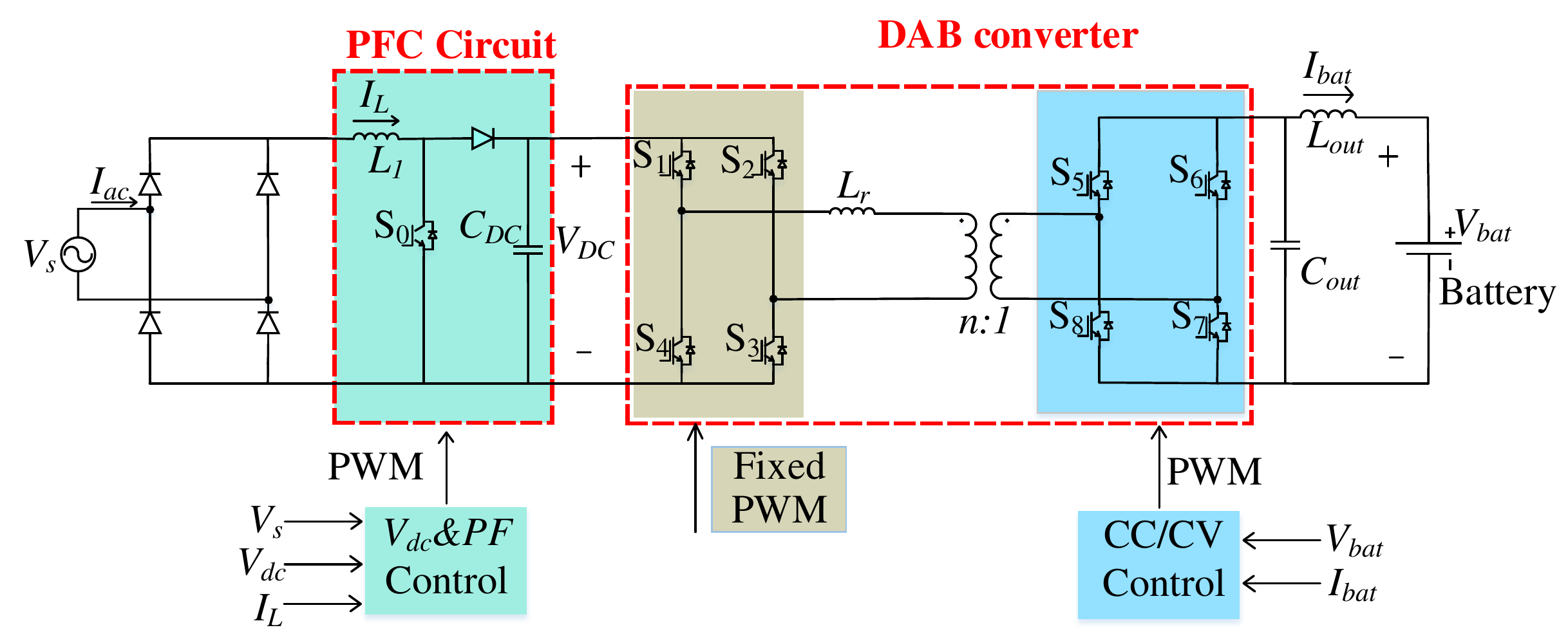}
\caption{\label{level1}Topology of a Level 1 or Level 2 charging system. $f_{PWM}=2$ $kHz$, $L_{out}=95$ $mH$, $C_{out}=0.1$ $mF$, $n=1$. For Level 1 model, $L_r=1$ $mH$, $V_s=120$ $V$ RMS. For Level 2 model, $L_r=0.13$ $mH$, $V_s=240$ $V$ RMS.}
\end{figure*}

Both Level 1 and Level 2 charging systems use single phase ac voltage as source.
The ac electricity firstly is passed to the diode-based AC/DC converter and is rectified to a DC
form. A PFC boost circuit is followed to adjust the DC voltage and improve the power factor. A
bidirectional DAB converter is used to convert dc voltage from one level to another level. The
CC/CV control is realized by the DAB converter control. A high frequency transformer is embedded in
the DAB for the galvanic isolation and to boost voltage. A typical Level 1 or Level 2 charging
circuit topology is shown in Fig. \ref{level1} as indicated in \cite{ReviewofBatteryCharger}. The
battery model, PFC and DAB control will be described and analyzed as follows.

\subsection{Battery model and charging control}
Battery is a key part in the EVs. The size of the battery determines the driving ranging and
charging time of the EVs. The battery stores chemical energy that converted from electricity and
releases electricity power to supply EVs working.

Currently, the lithium-ion battery is the most common and has been popularly used. Compared to
other types batteries, e.g., nickel-zinc or Ni-Cd battery, lithium-ion battery offers lighter
weight and higher power density. The capacity of a battery is measured as ampere-hours (Ah). The
stored energy is measured in watt-hours (Wh). The SOC is used to represent the current energy
available in battery. SOC for a fully charge battery is $100\%$ and $0\%$ for fully discharged.

 Assuming the battery terminal voltage is constant, the SOC can be defined as:
 \begin{equation}
    SOC=100\times (1-\frac{1}{C}\int_{0}^{t}i(t)dt)
\end{equation}
 where $C$ is battery capacity, $i(t)$ is battery terminal current.

\begin{figure}[ht!]
\centering
\includegraphics[width=2.8in]{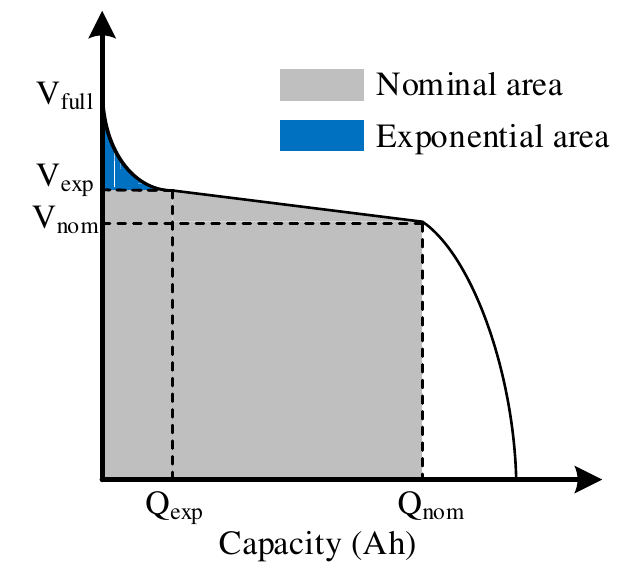}
\caption{\label{batchar}Battery charge and discharge curve. In this paper, 10 kWh battery is used. $V_{full}=291$ $V$, $V_{exp}=270$ $V$, $V_{full}=250$ $V$, $Q_{exp}=1.96$ $Ah$, $Q_{nom}=36.17$ $Ah$.}
\end{figure}

 In this paper, Matlab/SimPowerSystems' battery model is used. Fig. \ref{batchar} illustrates the charging and discharging process of a battery \cite{tremblay2009experimental}. A fully charged battery starts to discharge at full voltage. Then the voltage will keep an almost stationary value at the nominal area. This area releases most of stored energy. At the last region, the voltage drops
 rapidly.

The charging process is one of the most critical factors in battery application. Inappropriate
charging method may shorten battery life and even damage battery through overheating or
overcharging. Based on \cite{CCCVTC}, there are three types of charging methods: constant current
(CC), constant voltage (CV), and taper current (TC). CC charges the battery by keeping a constant
charging current. The charging process will stop when the voltage reaches a preset value. However,
to avoid overcharge, the current is required to set a small value. So this method is used in slow
charge. In contrast to CC, CV charging begins at a constant voltage with a decreasing current. This
method is usually used in less expensive EVs' chargers. However, since the voltage of each charging
battery are different, CV charging may generate high current and damage battery when the voltage
difference between battery and charger is significant. In TC charging, the charging current
decreases proportionally to the rising voltage. This method is often used in low capacity battery
and will not be considered in this paper.

\begin{figure}[h!]
\centering
\includegraphics[width=0.4\textwidth]{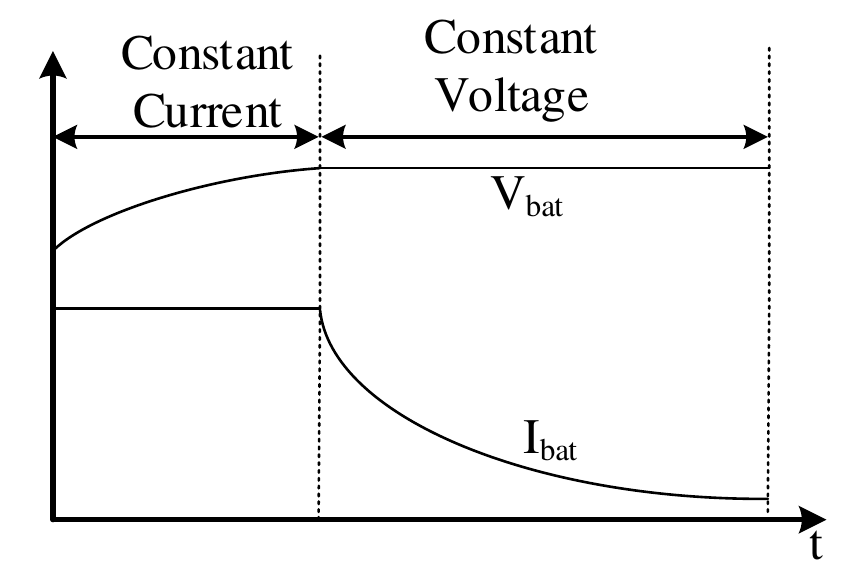}
\caption{\label{CCCVdiagram}Illustration of CC/CV control. The parameters of the model are: Level 1 $V_{CV}=$ $262$ $V$, $I_{CC}=$ $5$ $A$; Level 2 $V_{CV}=$ $273$ $V$, $I_{CC}=$ $40$ $A$; Level 3 $V_{CV}=$ $279$ $V$, $I_{CC}=$ $80$ $A$;}
\end{figure}

Based on aforementioned analysis, a charging method includes the integration of CC and CV charging is adopt in this project. The charging process is named as CC/CV and is shown in Fig. \ref{CCCVdiagram} \cite{ControlofaThree-Phase}. At the initial stage, the charging current is kept constant to avoid over-current. The charging will change to CV mode when the voltage reaches a preset value. At the CV mode, the battery is charged by a constant voltage with decreasing current. The most capacity of a battery is charged at CC mode, but the CV mode may take the same or longer time than CC.

\begin{figure}[ht!]
\centering
\includegraphics[width=0.50\textwidth]{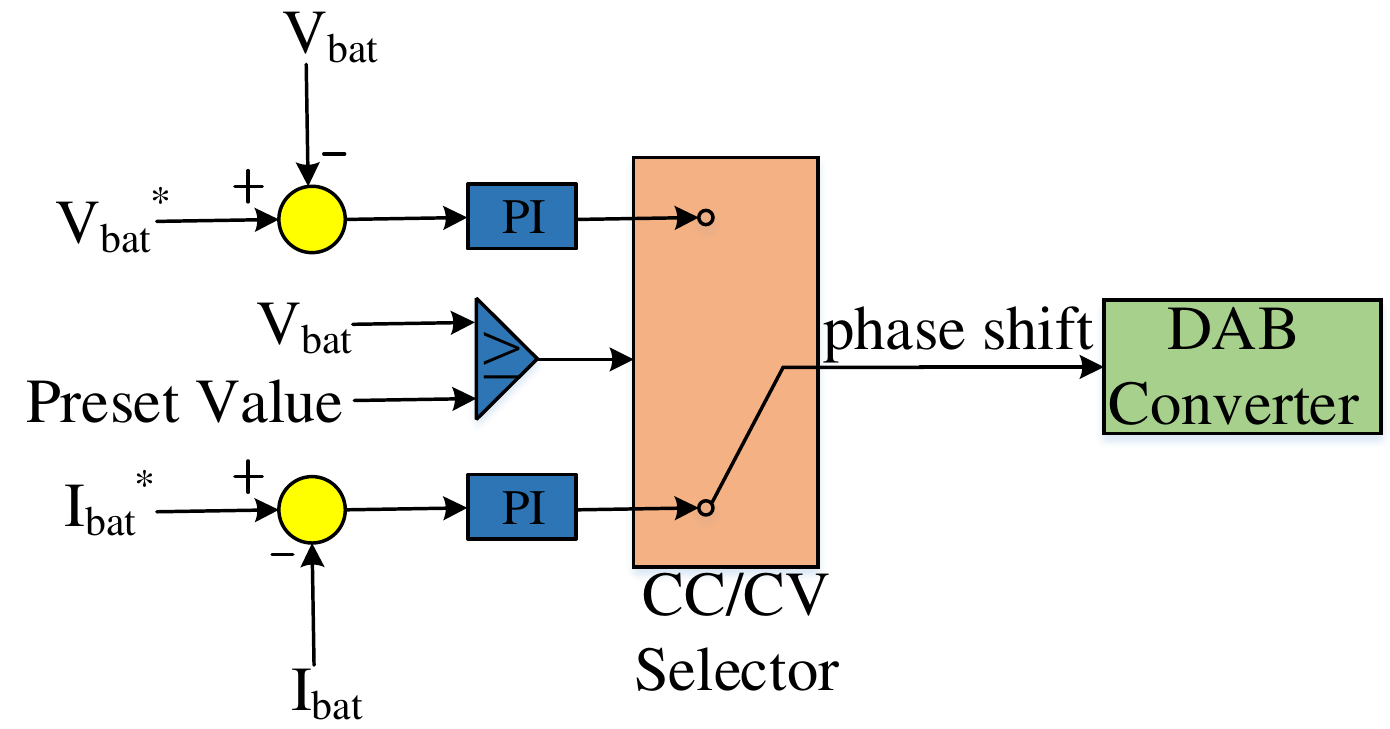}
\caption{\label{CCCV}CC/CV charging structure, the connection of selector depends on battery voltage. The preset value is constant voltage.}
\end{figure}

The implementation of CC/CV is shown in Fig. \ref{CCCV}. Two proportional integral (PI) controllers
are used to ensure current and voltage as constant, respectively. The function of the selector is
to switch CC or CV mode. In the beginning, the selector is connected to the current control mode to
realize CC control. The charging voltage is compared with a preset value. The selector will turn to
voltage control mode when the voltage reaches the preset value. The output of the control system is
phase shift. The phase shift is the phase shift between DAB's ac voltages of two ends. In subsection 2.2, how phase shift
can be realized and how it can influence power level will be explained.

\subsection{Dual active bridge converter}
Dual active bridge converter is widely used in the application of automobiles and renewable energy sources\cite{lehman2004,DABTS}. This paper adopts the DAB converter to fulfill the requirement of regulating battery charging current and voltage.

\begin{figure}[!h]
\centering
\includegraphics[width=0.6\textwidth]{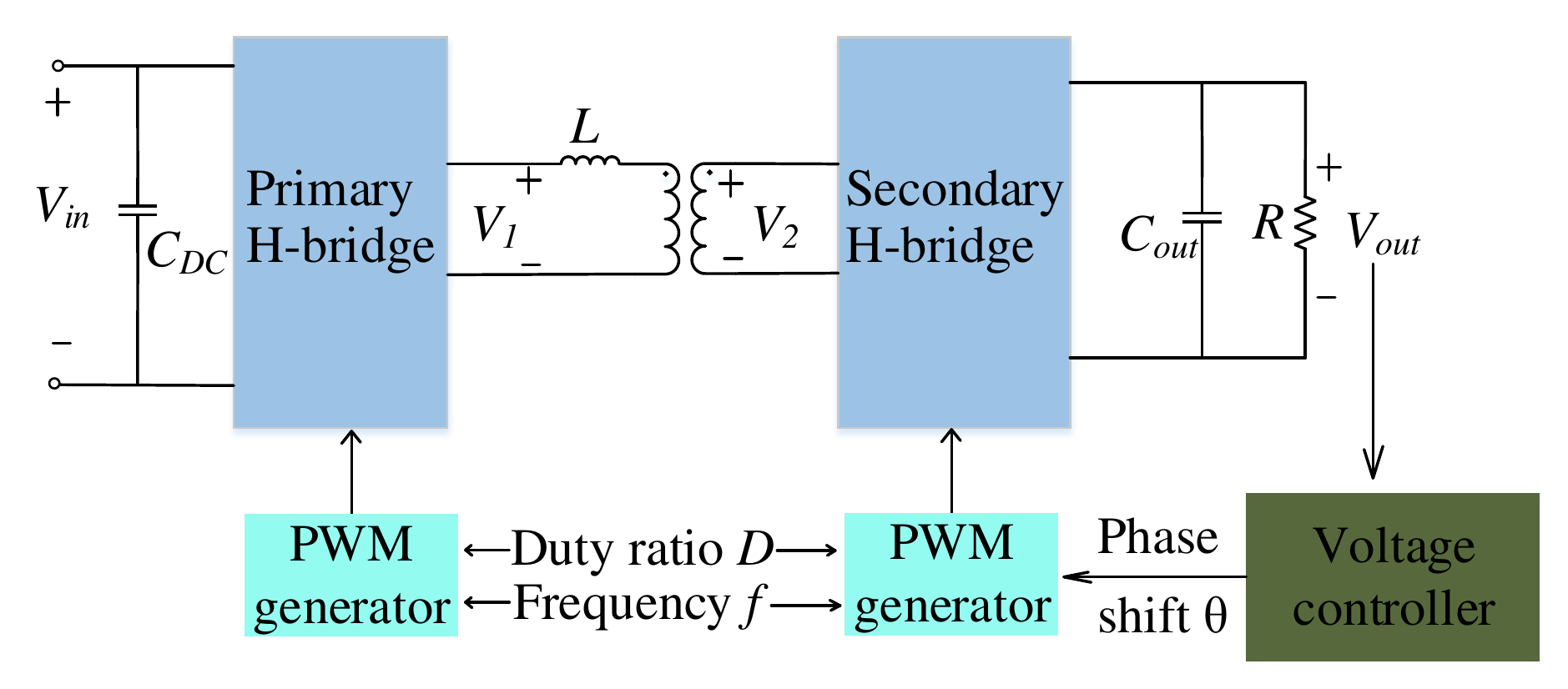}
\caption{\label{DAB}Structure of DAB converter, $f=2$ $kHz$, $D=0.5$.}
\end{figure}

\begin{figure}[!h]
\centering
\includegraphics[width=0.5\textwidth]{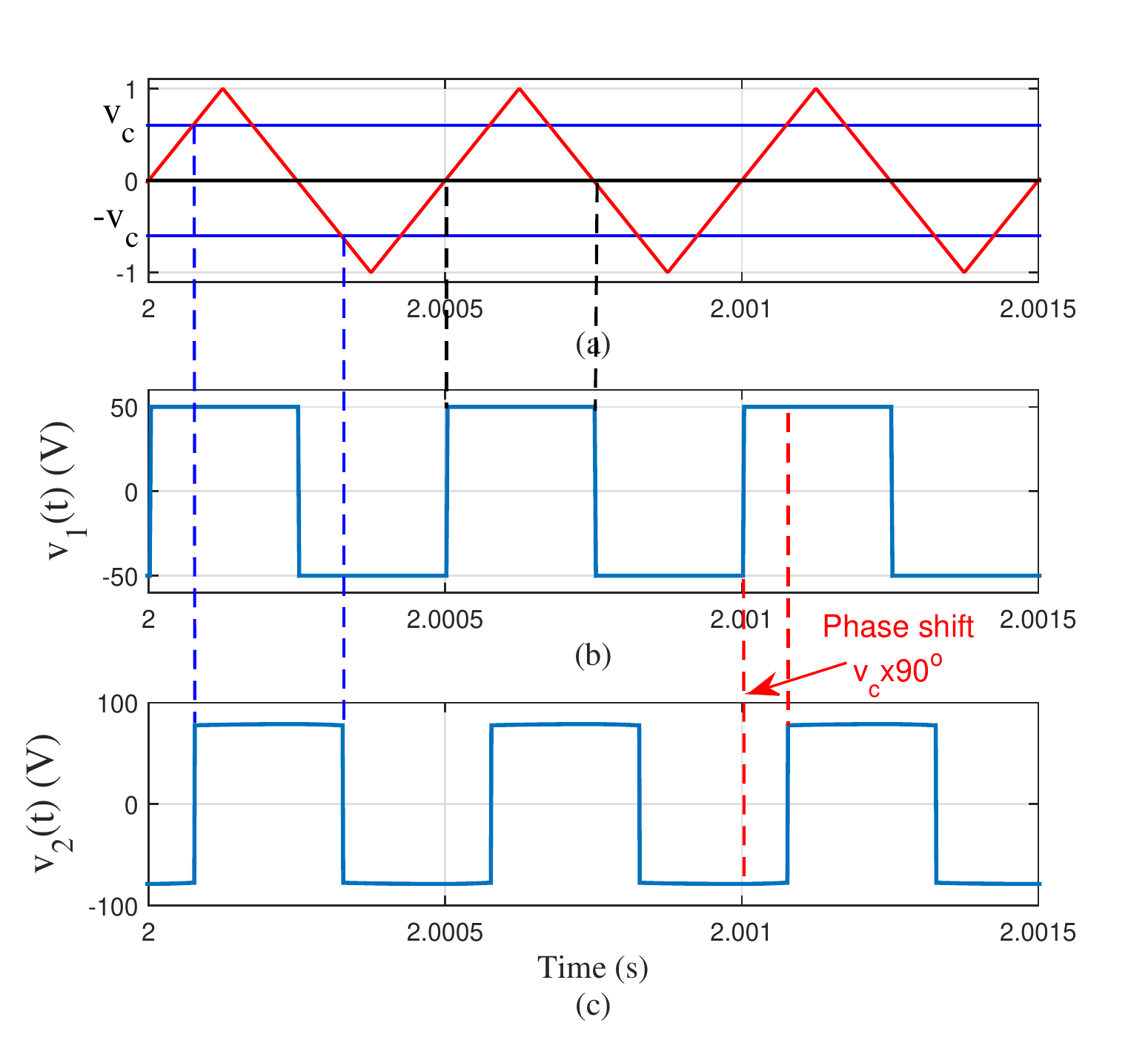}
\caption{\label{DABV1V2}DAB converter control waveform and voltages: (a) $v_c$, $-v_c$ and 0 compared with a triangle waveform, (b) primary side voltage, (c) secondary side voltage.}
\end{figure}

The DAB structure is shown in Fig. \ref{DAB}. The DAB consists of two H-bridge converters. A high-frequency transformer is used to connect the converters. Both the two converters have the same switching frequency and duty-ratio ($50\%$). The amount and direction of the power transfer can be changed by controlling the phase shift of the voltage on the two sides. The voltage phasors of the fundamental frequency on the two sides are denoted as $V_1$ and $V_2$. The phase shift of the two voltage is represented as angle $\theta$, which is shown in Fig. \ref{DABV1V2}.

The power transfer of the DAB can be expressed as:
\begin{eqnarray}
P=\frac{V_{1}V_{2}}{\omega L}\sin{\theta}
\end{eqnarray}
where $L$ is the inductor in the primary side, $\omega$ is related to the switching frequency as $2\pi\times2000$ $rad/s$.

Assuming the output capacitor $C_{out}$ ensures the output voltage $V_{out}$ be a constant, if we consider only the fundamental component, then the RMS values of $V_1$ and $V_2$ can be written as:
\begin{eqnarray}\notag
V_{1}&=&\frac{4}{\pi}\frac{1}{\sqrt{2}} V_{in} \\
V_{2}&=&\frac{4}{\pi}\frac{1}{\sqrt{2}} V_{out}   \notag
\end{eqnarray}

The power transfer can be expressed as (\ref{powtra}).
\begin{eqnarray}
\label{powtra}
P=\frac{V_{1}V_{2}}{\omega L}\sin{\theta}=\frac{\frac{1}{2}(\frac{4}{\pi})^2V_{in}V_{out}}{\omega L}\sin{\theta}
\end{eqnarray}

If all elements are ideal and there is no switching loss, the input power equals to the output power, shown in (\ref{PV}) and (\ref{PV1}).
\begin{eqnarray}
\vspace{-0.1in}
\label{PV}
\frac{\frac{1}{2}(\frac{4}{\pi})^2V_{in}V_{out}}{\omega L}\sin{\theta}=\frac{V_{out}^2}{R}\\
 \Rightarrow V_{out}=\frac{\frac{1}{2}(\frac{4}{\pi})^2RV_{in}}{\omega L}\sin{\theta}
 \label{PV1}
 \vspace{-0.1in}
\end{eqnarray}
where $R$ is assumed as the load resistance.

The small signal model of $V_{out}$ can be expressed as (\ref{linea}).

\begin{eqnarray}
\vspace{-0.1in}
\label{linea}
\Delta V_{out}=\underbrace{\frac{\frac{1}{2}(\frac{4}{\pi})^2RV_{in}\cos \theta}{\omega L}}_{K_1}\cdot \Delta\theta
\vspace{-0.1in}
\end{eqnarray}

Since $V_{in}$, $R$, $\omega$ and $L$ are all constants, then the output voltage $V_{out}$ is proportional to phase shift $\theta$.
\begin{eqnarray}
\vspace{-0.1in}
\Delta V_{out}	\propto \Delta \theta
\vspace{-0.1in}
\end{eqnarray}
Similarly, the current also can be expressed in small signal model as (\ref{Ilinea}).
\begin{eqnarray}
\vspace{-0.1in}
\label{Ilinea}
\Delta I_{out}=\underbrace{{\frac{\frac{1}{2}(\frac{4}{\pi})^2V_{in}\cos \theta}{\omega L}}}_{K_2}\cdot \Delta\theta
\vspace{-0.1in}
\end{eqnarray}

\begin{figure}[!h]
\centering
\includegraphics[width=0.45\textwidth]{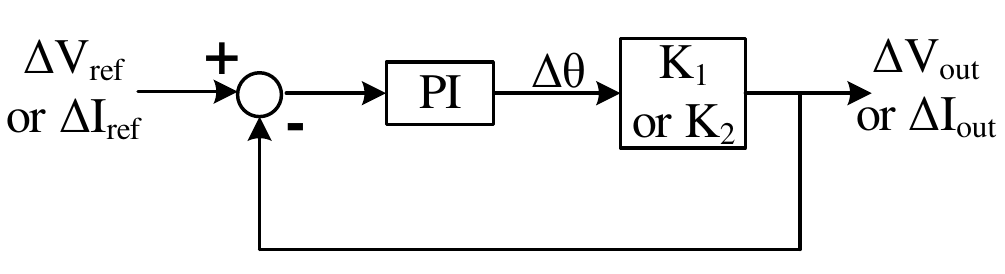}
    \caption{\label{DABPI}Diagram block of DAB converter. Input and output are in real value.}
\end{figure}

This equation shows the output voltage or current can be controlled by adjusting the phase shift
angle $\theta$. The block diagram of the phase shift control model is shown in Fig. \ref{DABPI}. To
make $V_{out}$ or $I_{out}$ follow a reference value, the error is fed into a PI controller to
generate phase shift. This phase shift can be realized by adjusting the control signal of the
receiving end converter PWM generator, as explained in Fig. \ref{DABV1V2}(a). If the reference
control signal $v_c$ is 0, the two blue lines will be aligned with the horizontal axis and 0 phase
shift generated. If the control signal value is 1, the two blue lines will be located at the top
and the bottom, which results in 90 degree phase shift. Phase shift is proportional to the control
signal $v_c$.

\begin{figure}[!ht]
    \centering
    \subfigure[]
    {
        \includegraphics[width=0.45\textwidth]{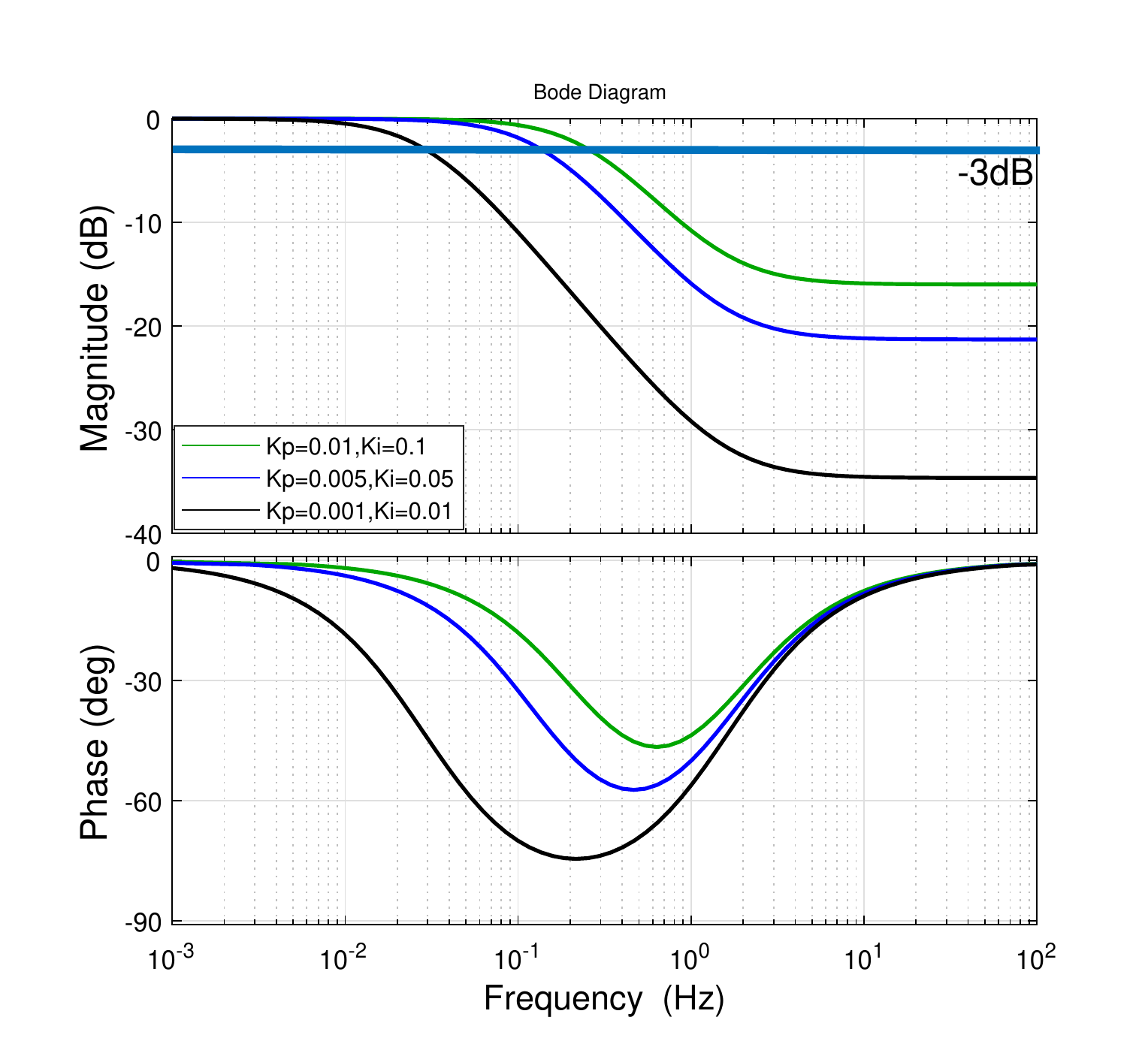}
    }
    \subfigure[]
    {
        \includegraphics[width=0.5\textwidth]{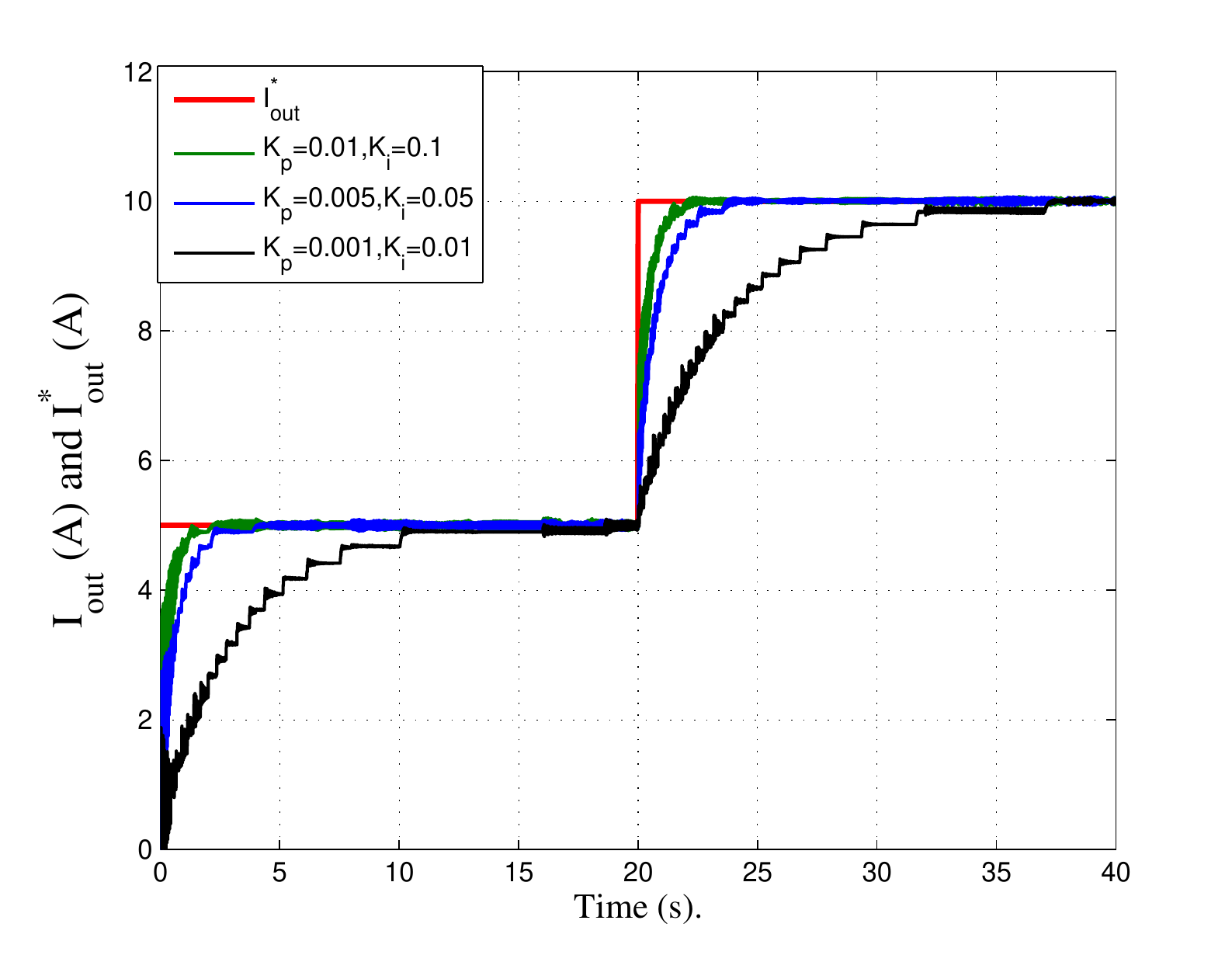}
    }
    \caption{The impact comparison of DAB PI controllers in Level 1 charger: (a) Bode plot, (b)step response. In Level 1 charger, $K_2=18.8$, when $k_p=0.01$ and $k_i=0.1$, the system has the largest bandwidth and fastest response time.}
    \label{DAB3Bode}
\end{figure}

The Bode plots and step response of the DAB closed-loop output current control model are illustrated in Fig. \ref{DAB3Bode}. Different PI values are applied to compare their dynamic characteristics. In this model, the PI value $0.01+\frac{0.1}{s}$ is chosen due to its fast dynamic performance.


\subsection{PFC boost circuit}
As Fig. \ref{level1} shows, a diode-based ac/dc converter is used to convert the ac voltage source to dc. However, the converter may generate non-sinusoidal currents thus results in low efficiency. Thus, a power factor correction (PFC) boost circuit is employed to achieve higher power factor or even unity power factor.

\begin{figure}[!h]
\centering
\includegraphics[width=0.55\textwidth]{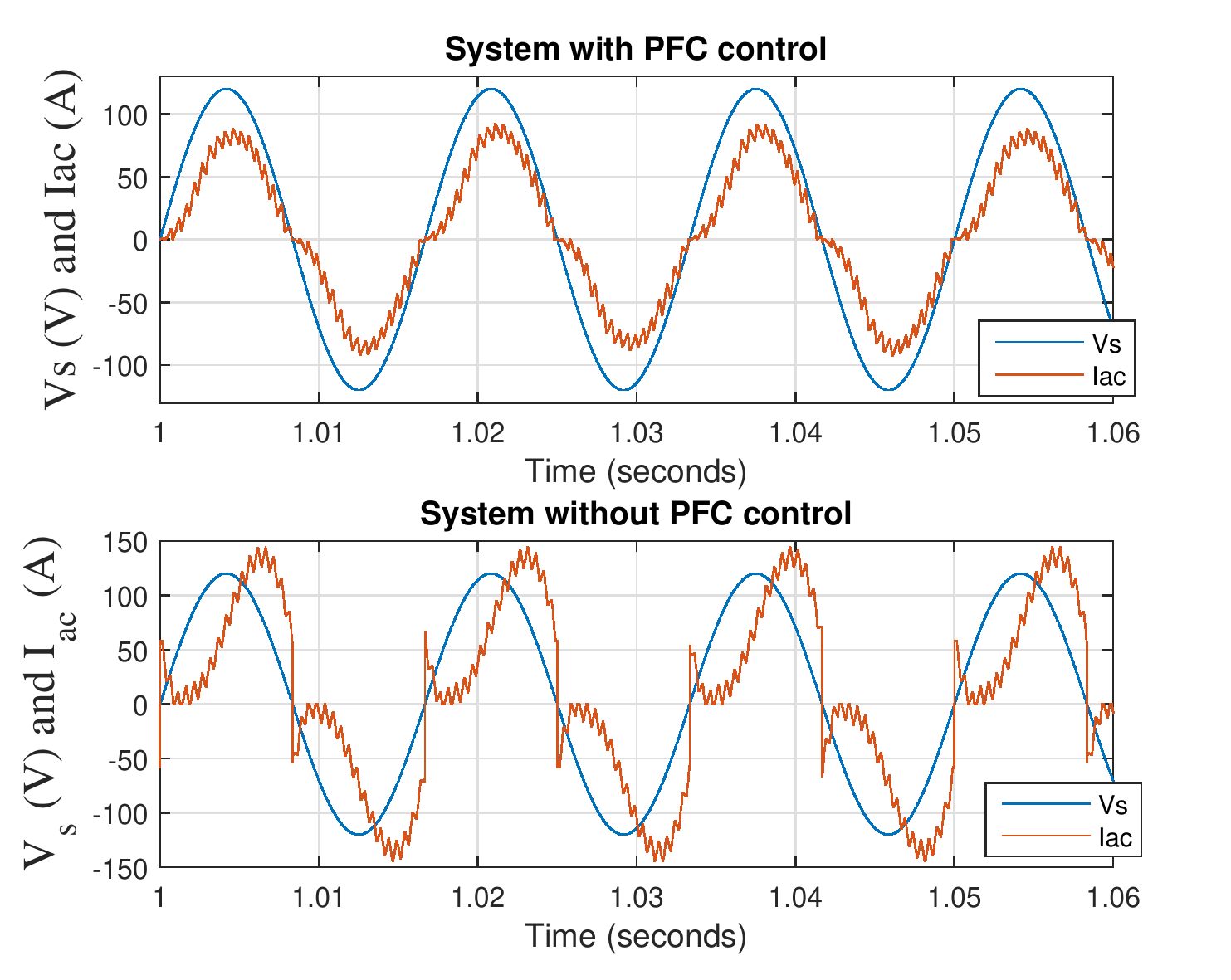}
\caption{\label{PFCIV}Comparison of boost converter performance with and without PFC control. The system without PFC control has a fixed duty ratio.}
\end{figure}

The PFC improves power factor (PF) by reshaping input current to be in phase with the input voltage\cite{PFC, PFC1}. Fig. \ref{PFCIV} demonstrates the impact of PFC by comparing the input voltage and current in the system with and without PFC. The current in the system with PFC is almost a sinusoidal waveform and has the same phase with the input voltage. In contrast, without PFC, the current shows the harmonics and phase shift with voltage.

\begin{figure}[!h]
\vspace{-0.1in}
\centering
\includegraphics[width=0.45\textwidth]{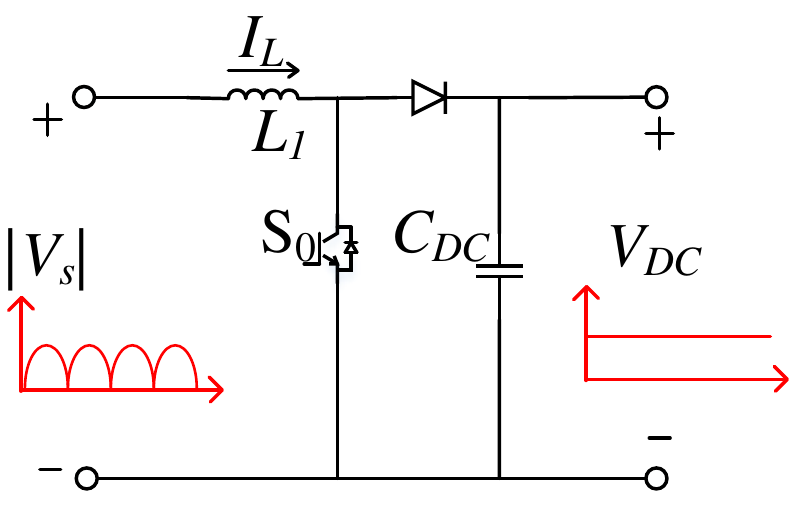}
\caption{\label{PFCblock}PFC circuit.}
\vspace{-0.1in}
\end{figure}

The PFC circuit is shown in Fig. \ref{PFCblock}, the circuit includes two vital parameters: DC bus capacitor $C_{\rm DC}$ and inductor $L_1$. The two parameters determine the performance of the circuit. Thus, an appropriate design for the parameters is necessary. The values are deigned based on following analysis.

Assuming that the PFC has a unity power factor, the instantaneous active power of the charger is as
follows.
\begin{align}
p(t)  = P + P\cos(2\omega t)
\end{align}
where $\omega$ is 377 rad/s and $P$ is the active power.

The output capacitor ensures that the output voltage can be treated as a constant dc voltage. The
active power is consumed by load and the second frequency power ripple is expected to be absorbed
by the capacitor. The output current is $I_{\rm out}=\frac{P}{V_{\rm DC}}$, and the capacitor
current is calculated as (\ref{ic}).
\begin{align}
\label{ic}
i_c(t)  = \frac{P\cos(2\omega t)}{V_{\rm DC}} = I_{\rm out}\cos(2\omega t) .
\end{align}

Since the current through a capacitor is determined by $C\frac{dv_c}{dt}$, the voltage ripple can be found by integrating capacitor current as (\ref{vc2}) shown.
\begin{align}
\label{vc2}
v_{c2}(t)  = \frac{1}{C_{\rm DC}}\int i_c(t)dt =\frac{I_{\rm out}}{2\omega C_{\rm DC}}\sin(2\omega t).
\end{align}

Thus the peak-to-peak voltage ripple is presented by (\ref{vc}).
\begin{eqnarray}
\label{vc}
\Delta V_C=\frac{I_{\rm out}}{2\pi fC_{\rm DC}}
\end{eqnarray}

The battery voltage ranges $200$ $V$ to $400$ $V$ so the PFC circuit should support the maximum
output voltage, $400$ $V$\cite{krein1998elements}. According to Table. \ref{tab1}, the maximum
power rating is $14.4$ kW. So considering the maximum output voltage, the current is $\frac{14.4\>
kW}{400 \>\text{V}}=36$ $A$. If the peak-to-peak voltage ripple is less than $10$ $V$, and the
current is chosen as $36$ A, then the capacitor is calculated as $9.5$ $mF$.

The inductor current ripple can be found based on (\ref{vl}).
\begin{eqnarray}
\label{vl}
V_L=L\frac{di}{dt}\Rightarrow V_{\rm out}-V_{\rm in}=L\frac{\Delta i_L}{DT}
\end{eqnarray}
where D is duty ratio and T is switching period.

Since the PFC boost circuit is able to handle a wide range of output voltage, then the output is
chosen as maximum value, 400 $V$, and the input is Level 1 input voltage, $120$ $V$ RMS single
phase voltage. For a 120 $V$ AC voltage, the mean value of the rectified voltage is 108 $V$. So the
duty ratio is $D=\frac{V_{\rm out}-V_{\rm in}}{V_{\rm out}}=0.73$. The circuit is operated at a
switching frequency of 2 $kHz$, and if 1 $A$ current ripple is allowed, then the inductor needs to
be 98 $mH$.

\begin{figure}[!h]
\centering
\includegraphics[width=0.55\textwidth]{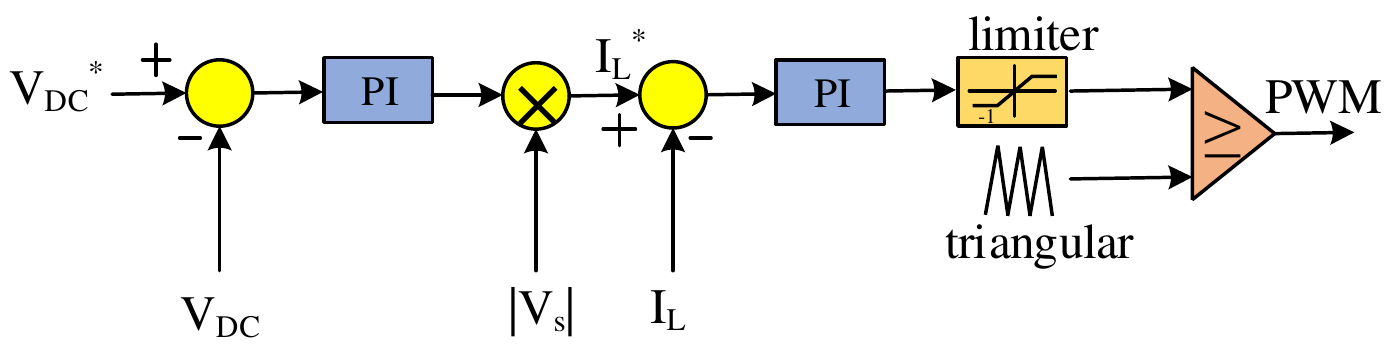}
\caption{\label{PFC}Block diagram of PFC control. The outer loop controller is $0.0005+\frac{0.005}{s}$, inner loop controller is $0.01+\frac{0.1}{s}$.}
\end{figure}

The PFC controller consists of two control loops: inner inductor current loop and outer DC bus
voltage loop. Fig. \ref{PFC} shows a PFC controller block \cite{PFC}. The outer PI controller is
utilized to ensure the DC bus voltage $V_{DC}$ of the boost converter follow a fixed reference
voltage. The output of the $V_{DC}$ PI controller is multiplied by the rectified sinusoidal
voltage to generate a reference inductor current which is in the same phase with the rectified
voltage. Similarly, the inner PI controller regulates the inner inductor current to follow the
reference current. The inner PI controller generates the duty ratio of the PFC boost converter.
Thus, the PFC controller is able to shape the current to be synchronized with voltage and regulate
$V_{DC}$.

\begin{figure}[!h]
\centering
\includegraphics[width=0.6\textwidth]{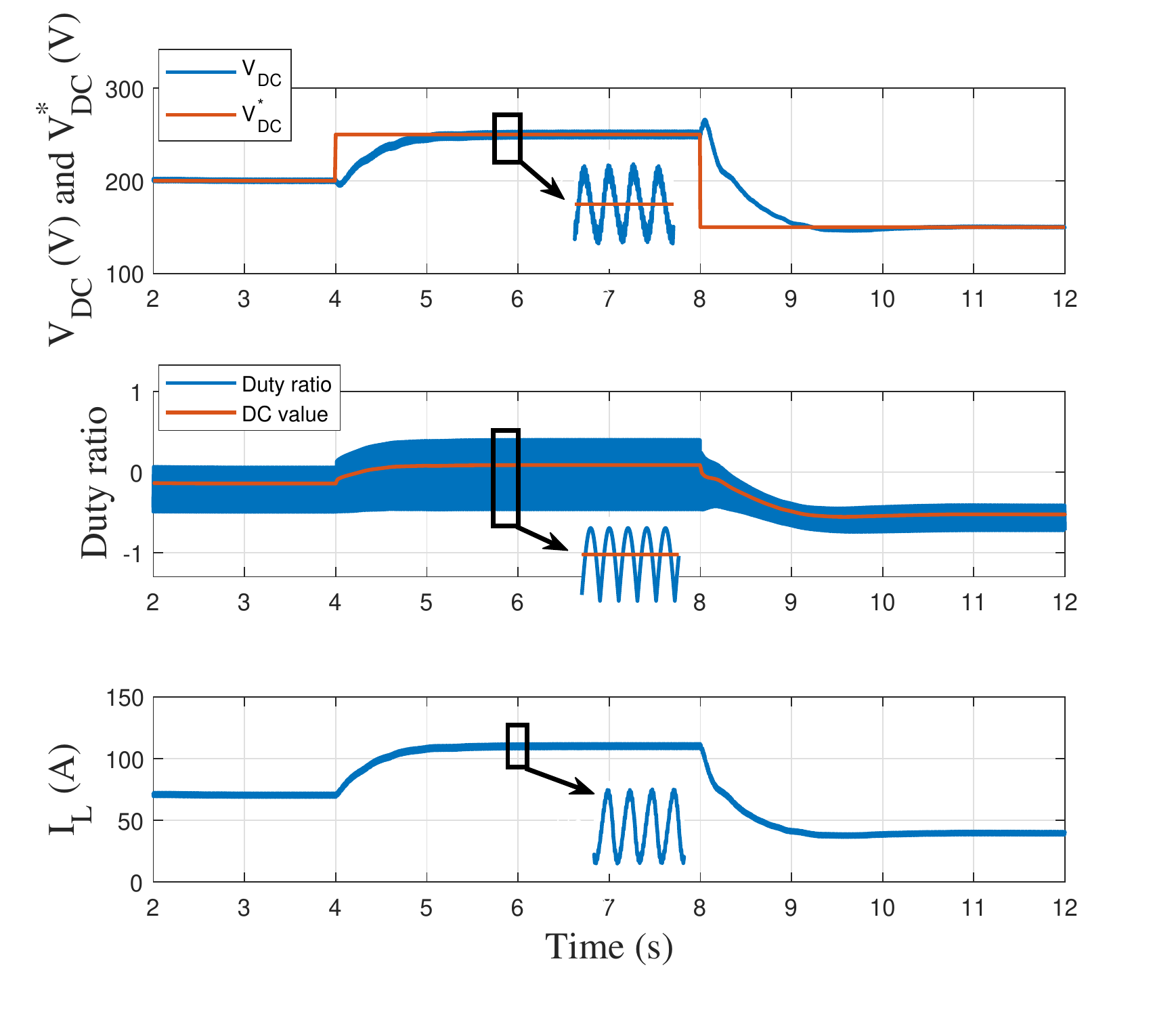}
\caption{\label{Vandd}Step response of a PFC example.}
\end{figure}

Fig. \ref{Vandd} shows the dynamic performance of this PFC. A step change to the PFC is applied by increasing the reference output voltage at $t=4s$ from $200$ $V$ to $250$ $V$ and decreasing it at $t=8s$ to $150$ $V$. It can be observed that the duty ratio and inductor current also have an increase and decrease followed by reference voltage.

\section{Level 3 charging circuit topology}
\begin{figure}[!h]
\centering
\includegraphics[width=0.9\textwidth]{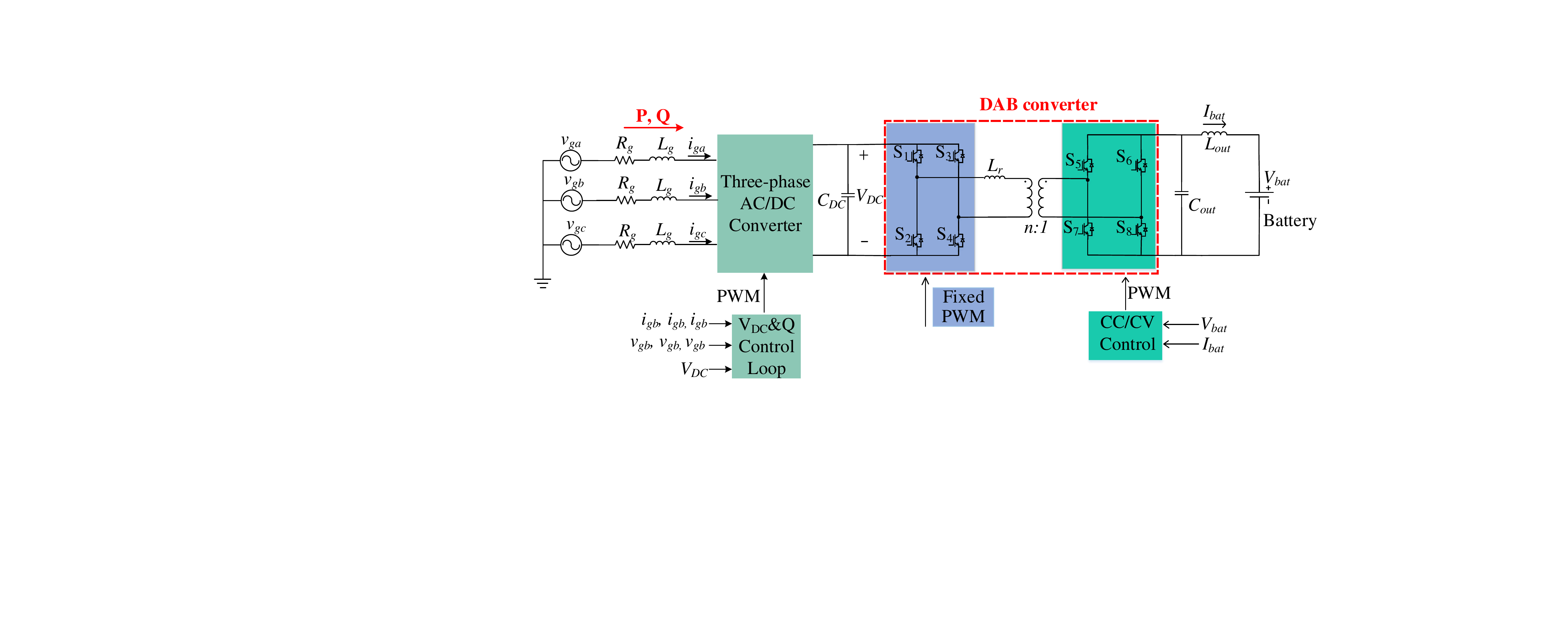}
\caption{\label{level3}Topology of level-3 charging system, $R_g=3$ $m\Omega$, $L_g=3$ $mH$, $C_{DC}=30$ $mF$, $L_r=0.13$ $mH$, $f_{PWM}=2$ $kHz$, $L_{out}=100$ $mH$, $C_{out}=0.1$ $mF$, $n=1$.}
\end{figure}

Since the Level 3 charging system uses three-phase voltage as power source, a three-phase converter
is required. In this project, a bi-directional voltage source converter is adopted. This converter is
controlled by a $V_{DC}/Q$ controller which regulates the DC bus voltage and reactive
power that receives from grid. A DAB bidirectional converter is followed. The DC bus voltage is the
input of the DAB converter, which controls the battery charging voltage or current using CC/CV. In
addition, this charger is able to realize the vehicle to grid (V2G) service by controlling the
power flow direction. The topology of Level 3 charger is shown in Fig. \ref{level3}.

\subsection{$V_{DC}$/Q controller}
\begin{figure}[!h]
\centering
\includegraphics[width=0.65\textwidth]{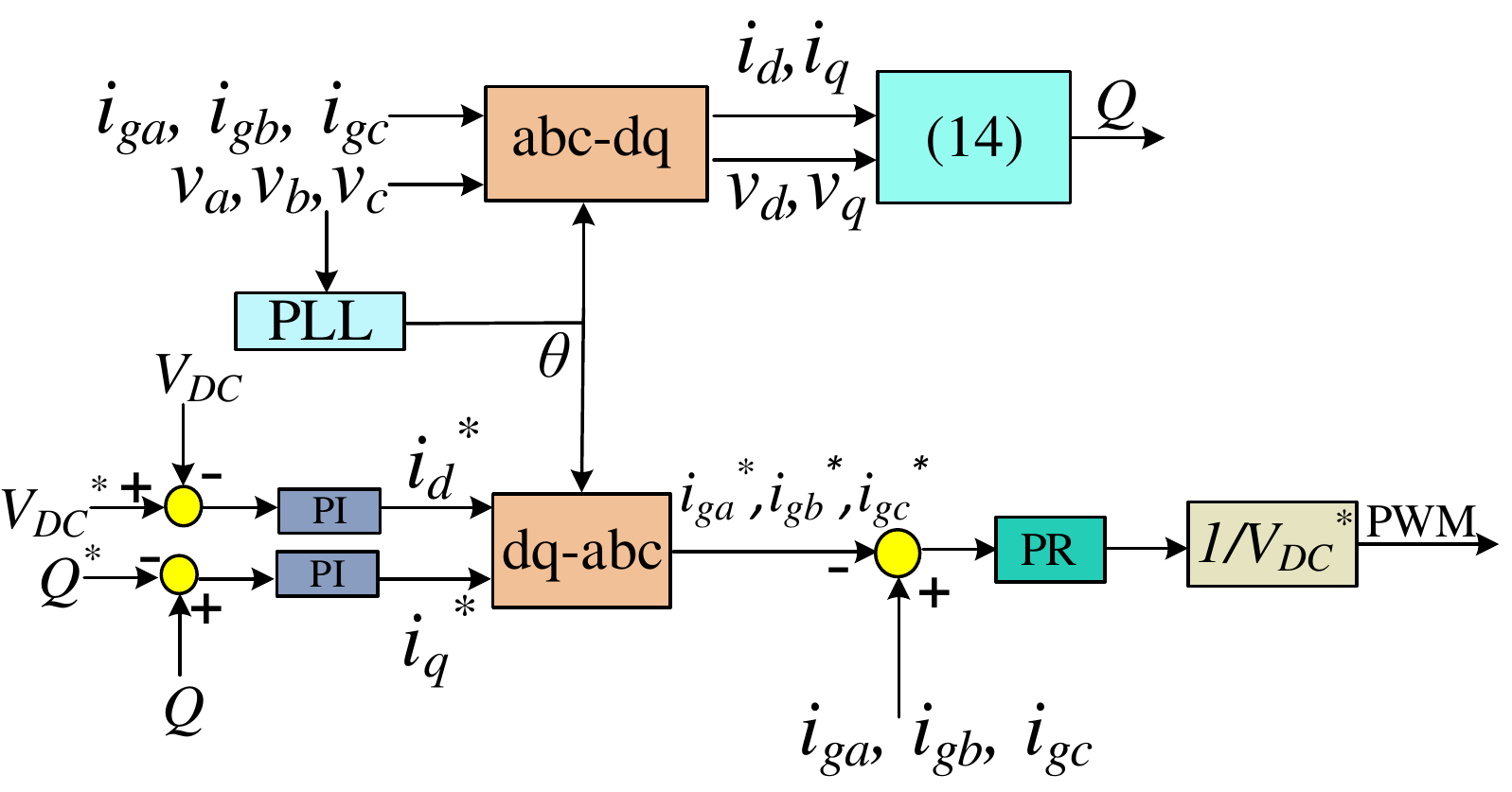}
\caption{\label{VdcQ}$V_{\rm DC}/Q$ control method applied to the three-phase converter, the $V_{DC}$ PI controller is $0.1+\frac{1}{s}$, reactive power controller is $0.01+\frac{0.1}{s}$.}
\label{control_level3}
\end{figure}

The $V_{DC}/Q$ controller structure is shown in Fig. \ref{VdcQ}. The controller receives the
measurements of DC bus voltage $V_{DC}$ and reactive power $Q$ from the grid. The two measurements
are compared with their reference values. The differences are fed into two PI controllers and
results in dq-axis current orders, which are converted into three-phase form by use of the angle of
the input voltage space vector ($\theta$). The grid voltage $v_a$, $v_b$ and $v_c$ are injected
into a phase-locked-loop (PLL) to generate the angle ($\theta$). The three-phase input current and
voltage are converted into $dq$-variables based on a $dq$-frame with its d-axis aligned to the grid
voltage. Real and reactive power absorbed by the EV from the grid are expressed as follows.
\begin{eqnarray}
\label{Q}
P&=&\frac{3}{2}(v_d i_d+v_q i_q)=\frac{3}{2}v_d i_d,\\
Q&=&\frac{3}{2}(v_q i_d-v_d i_q)=-\frac{3}{2}v_d i_q.
\label{P}
\end{eqnarray}
Note that the grid voltage is aligned to the $d$-axis. Hence $v_q=0$. In addition, $P$ can be
adjusted by varying $i_d$ while $Q$ can be adjusted by varying $i_q$. Due to the negative linear
relationship, reactive power control will be a positive feedback control, as shown in Fig.
\ref{control_level3}.

Three proportional and resonant (PR) controllers are utilized to control the three-phase converter. The input of the PR controller is the error between measured input current and its reference which is generated from dq/abc conversion. The PR controller generates converter voltage, which will be scaled to generate PWM waveform applied to the three-phase converter.

\subsection{PR controller}
Three PR controllers are used in the $V_{DC}/Q$ control system. The function of a PR controller is to eliminate steady-state error at a designed frequency. In contrast, the PI controller can only realize the function at zero frequency.

The transfer function of the PR controller is shown in :
\begin{eqnarray}
G(s)=K_p+K_r\frac{2\omega_cs}{s^2+2\omega_cs+\omega^2}
\end{eqnarray}
where $\omega$ represents resonance frequency, $\omega_c$ is cutoff frequency, $K_p$ is the proportional gain and $K_r$ represents the resonant gain.

\begin{figure}[!ht]
\centering
\includegraphics[width=0.65\textwidth]{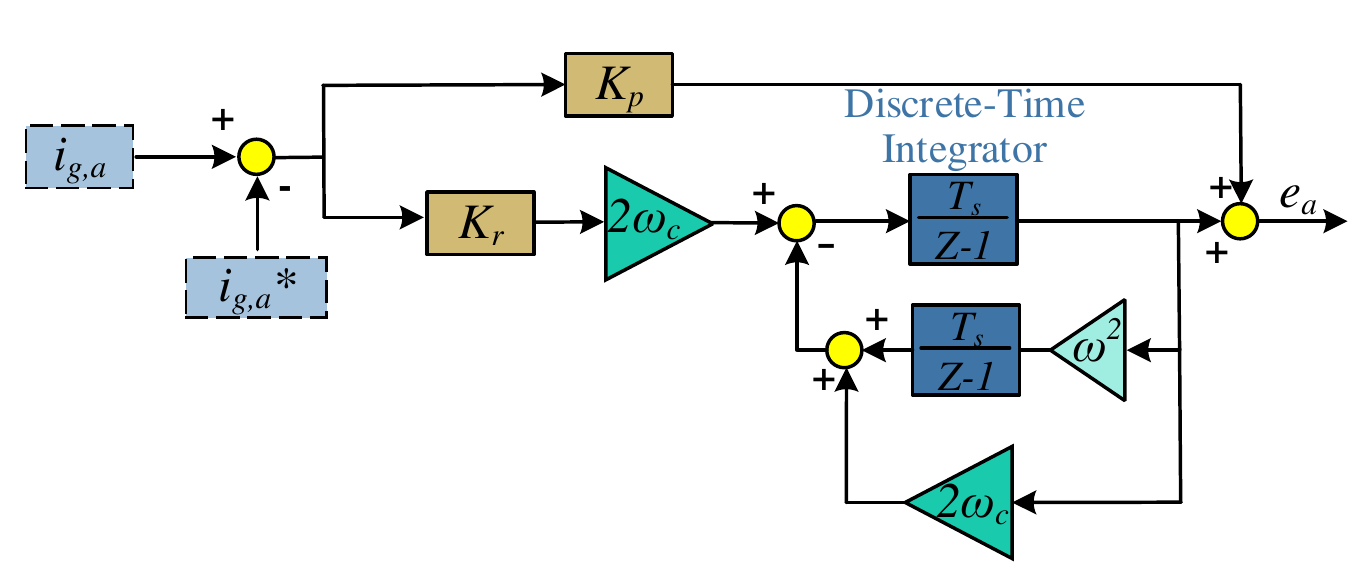}
\caption{\label{PR}Structure of PR controller in phase a.}
\end{figure}

Note that the RT-Lab runs at discrete step size, so the PR controller should be modified to
discrete time domain. Thus, the integrator $\frac{1}{s}$ is replaced by $\frac{T_s}{z-1}$, where
$T_s$ is the simulation step size, e.g., 50 $\mu$s. The block diagram of a discrete PR controller
is shown in Fig. \ref{PR}.

\begin{figure}[!ht]
\vspace{-0.15in}
\centering
\includegraphics[width=0.55\textwidth]{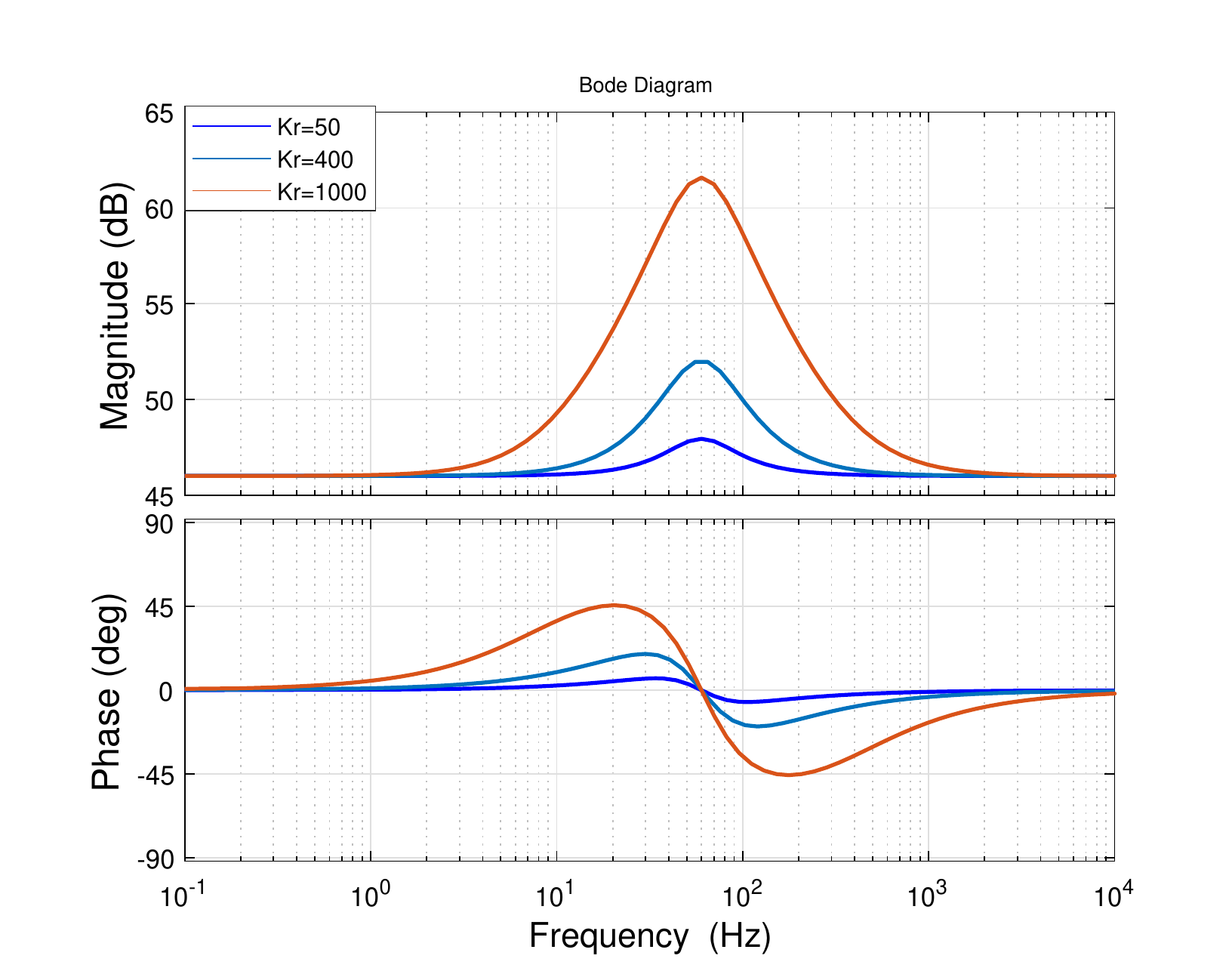}
\caption{\label{PRBode}Bode plot of PR controller with varies $K_r$ and $K_p=200$, $\omega_c=200$, $\omega=2\pi \times 60$.}
\vspace{-0.15in}
\end{figure}

Considering the resonance frequency as $377$  $rad/s$, both $K_p$ and $\omega_c$ are fixed at 200. The Bode plot with
varies $K_r$ is shown in Fig. \ref{PRBode}. As the Bode plot shows, the block gain is very large at
the resonance frequency is $377$ $rad/s$. So the PR controller is able to track a signal that
frequency is $377$ $rad/s$ and the steady-state error will be eliminated. The $K_r$ determines the
width of the frequency bandwidth and dynamics. As illustrated in this Bode plot, a high $K_r$
achieves a wider bandwidth compared to a low $K_r$. Thus, a high value is selected for $K_r$. In
this model, PR controller is chosen as $200+1000\frac{400s}{s^2+400s+377^2}$.

\subsection{Phase locked loop}
\begin{figure}[!ht]
\vspace{-0.07in}
\centering
\includegraphics[width=0.6\textwidth]{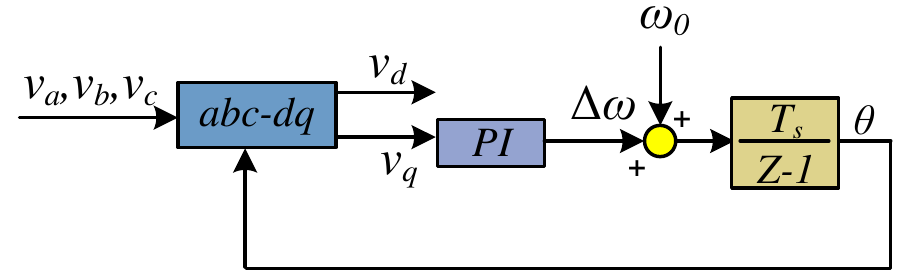}
\caption{\label{PLL}Block diagram of a PLL, the PI controller is $0.05+\frac{1}{s}$. The input voltages are in real values.}
\vspace{-0.15in}
\end{figure}

The function of a phase-locked-loop (PLL) is to synchronize the charging system to the grid.
Three-phase grid voltage $v_a$, $v_b$, $v_c$ are input to the PLL and then converted into dq-axis
variables by using of the angle $\theta$ from PLL. The angle is the grid voltage space vector angle
measured by PLL. A second-order PLL is adopted in this project, the diagram block is shown in Fig.
\ref{PLL} \cite{fan2017control}. A PI controller is used to ensure $v_q$ zero. The reference grid
current $i_d^*$ and $i_q^*$ generated by DC bus voltage and reactive power will be converted to
a-b-c form by using of the angle $\theta$ from PLL.

\begin{figure}[!h]
\vspace{-0.15in}
\centering
\includegraphics[width=0.75\textwidth]{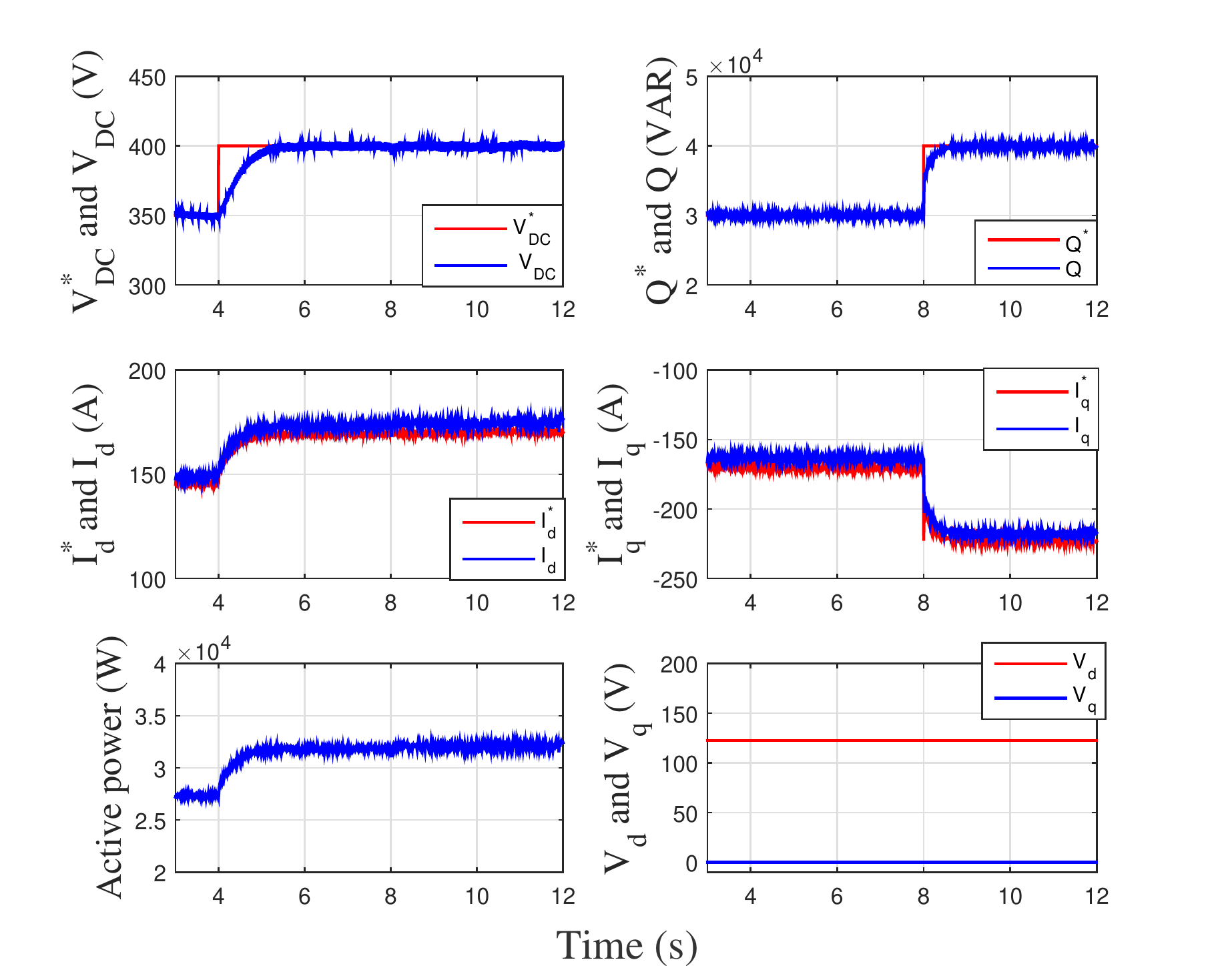}
\caption{\label{VDCQ}Step response of $V_{DC}$/Q control}
\vspace{-0.15in}
\end{figure}

A step change is applied to the controller as shown in Fig. \ref{VDCQ}. At $t=4s$ DC bus reference voltage increases from $350$ $V$ to $400$ $V$. At $t=8s$ reactive power increases from $30$ $\rm kVAR$ to $40$ $\rm kVAR$. The result shows that the DC bus voltage and reactive power can be well controlled.

\section{Structure and performance of RT-Lab simulator}

Considering the complicated model and long simulation time, real-time simulator RT-Lab is adopted.
Firstly, the models are built in Matlab/SimPowerSystems. Then the models are converted to C code
and implemented in RT-Lab. The RT-Lab simulates the power system model at the same rate as the
physical time. Thus it offers much faster simulation speed.

A RT-Lab model is separated into three subsystems as master, slave and console subsystem. The master subsystem includes the computational and control part \cite{ORT}. For example, the PFC control, CC/CV control, and $V_{DC}/Q$ control are assigned to the master subsystem. The slave subsystem consists of the circuit elements that include the voltage source, battery model, IGBTs, etc. The master and slave subsystem are located on different CPUs of the RT-Lab, so they are implemented in parallel to increase computation speed. The console subsystem is used to monitor the model while the model is executing. Users also can change the parameters of the model during simulation. It has to be mentioned that the console subsystem does not need devoted CPU.

\begin{figure}[!ht]
\centering
\includegraphics[width=0.55\textwidth]{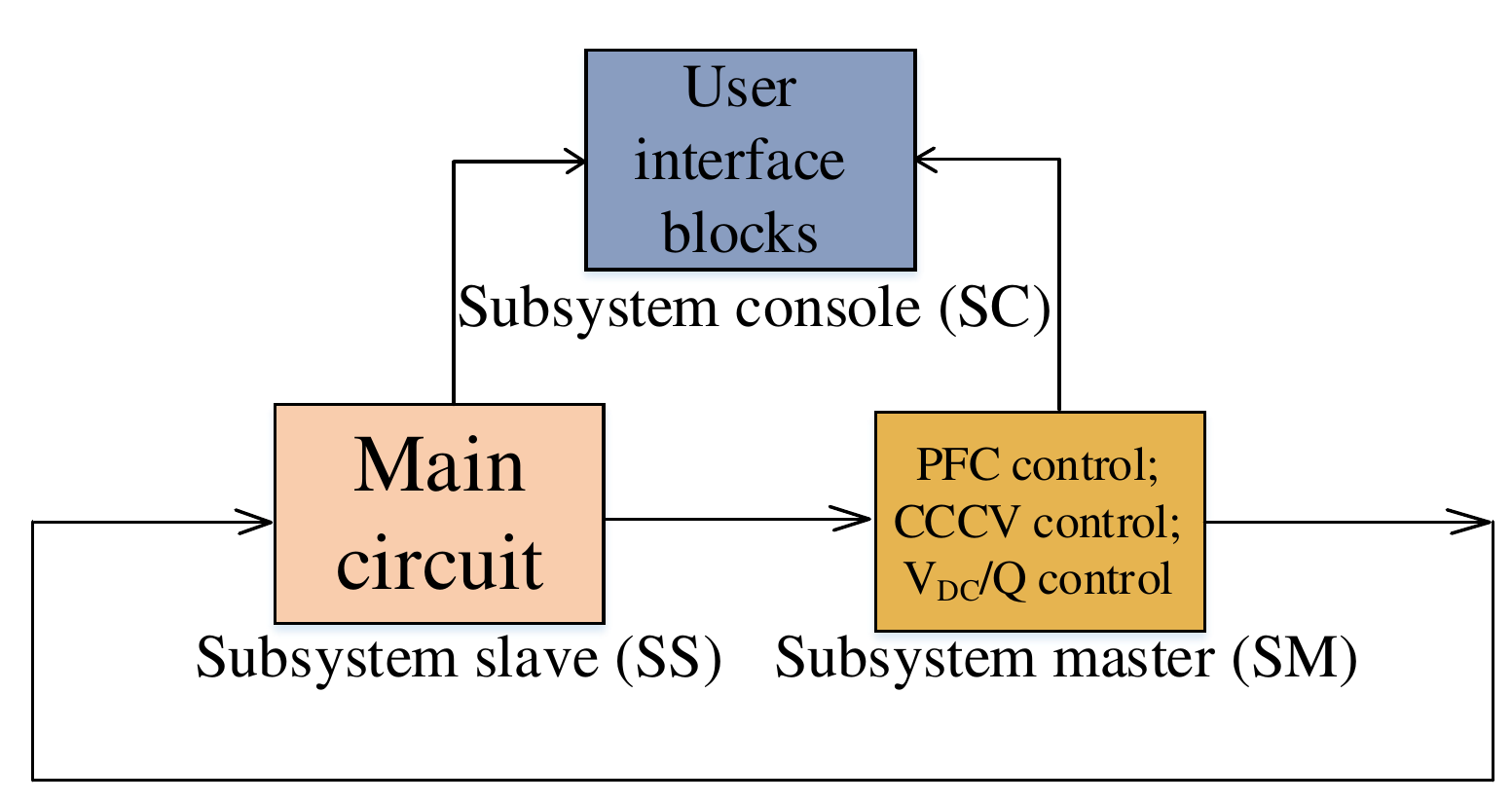}
\caption{\label{RTstru}RT-Lab model structure.}
\end{figure}

The charging system real-time simulation modeling structure is shown in Fig. \ref{RTstru}. The
slave system includes the main circuit, where the measured parameters such as battery voltage and
current, DC bus voltage are extracted and put into master subsystem. In the master subsystem, these
parameters will be analyzed and control signals will be generated and sent to the main circuit.

\begin{figure}[!ht]
\vspace{-0.15in}
\centering
\includegraphics[width=0.75\textwidth]{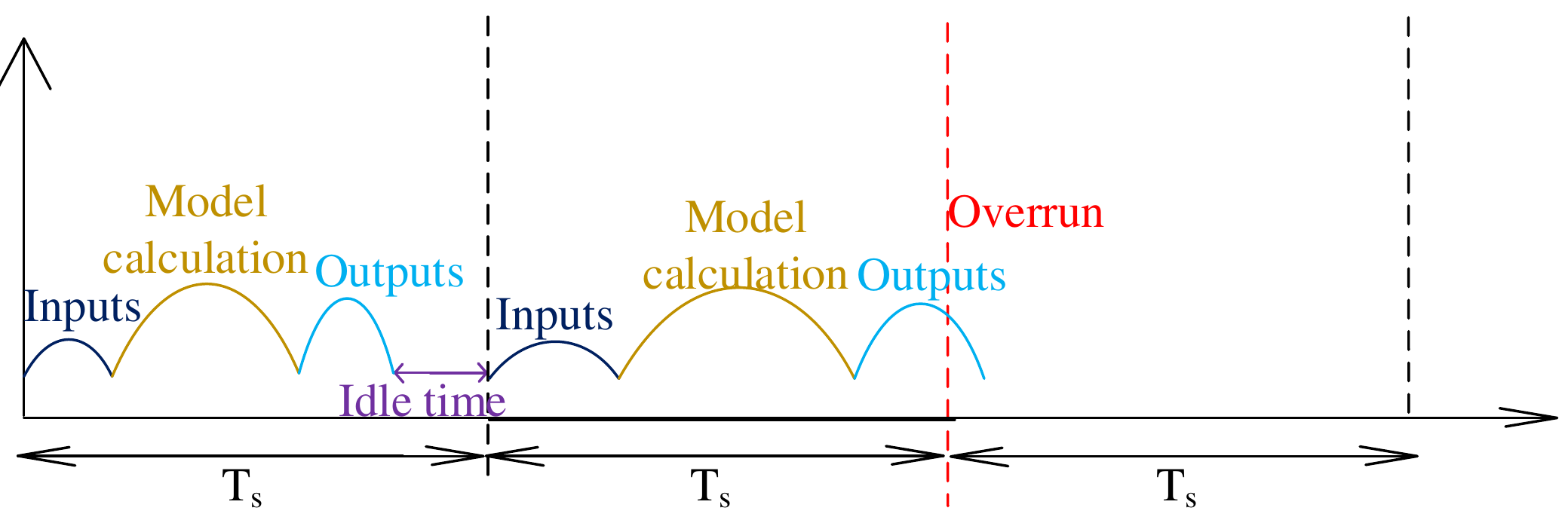}
\caption{\label{stepsize}RT-Lab model time-step structure.}
\vspace{-0.15in}
\end{figure}

In each simulate time step, the RT-Lab needs to complete three procedures: input, calculation and output. At the beginning, the RT-Lab reads the input data and then copes with these data based on functions. After processing these data, the results will be sent back to the model. The execution time of the three procedures should be less than the step size. Otherwise an error 'overrun' will happen in the model. The 'overrun' may cause the next step time omitting and thus data lost \cite{RTS}. So the overrun should be avoided to make sure the accuracy of the model. If the three procedures take less time than the step size, the rest time is called idle time. The process of the simulation is shown in Fig. \ref{stepsize}.

\section{Simulation results}
This section shows the real-time simulation results of proposed three-level charging systems in
RT-Lab simulator. The battery used in the system has the power rating of $10$ $kWh$ and capacity is
$40$ $Ah$. The overall simulation shows the charging voltage, current, power and phase shift, DC
bus voltage, PFC inductor current, reactive power and battery SOC comparison.

\begin{figure}[!ht]
\centering
\includegraphics[width=1.1\textwidth]{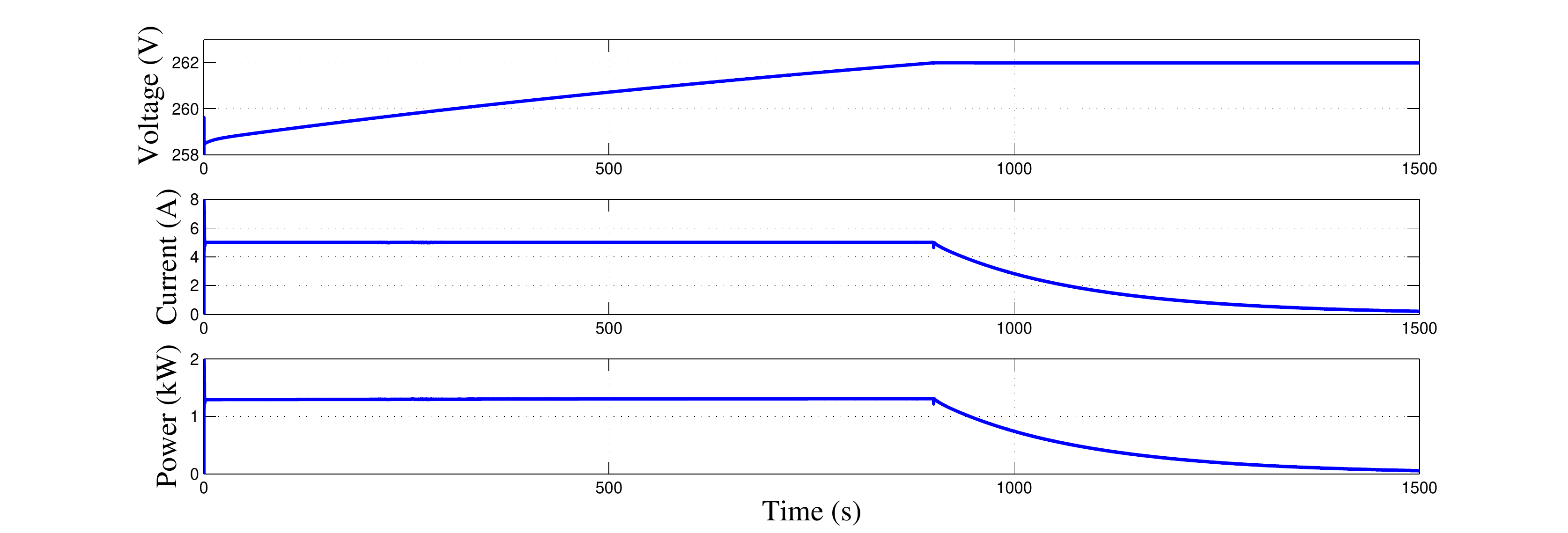}
\caption{\label{12}Current, voltage and charging power of Level 1 charging.}
\end{figure}

\begin{figure}[!ht]
\centering
\includegraphics[width=1.1\textwidth]{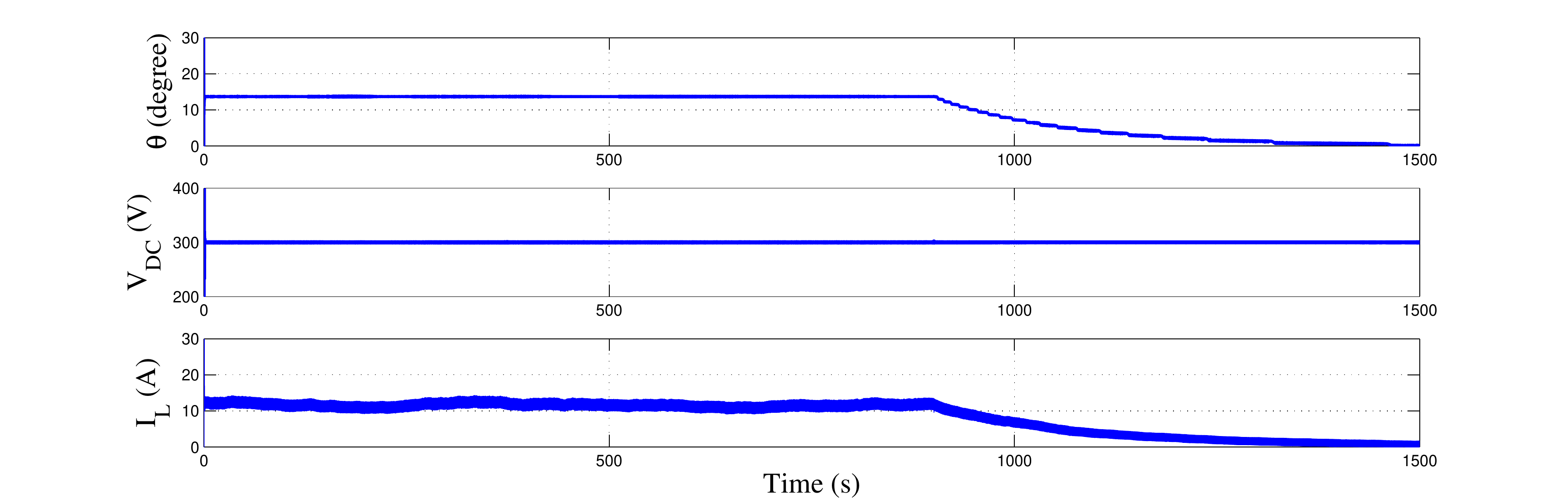}
\caption{\label{13}Phase shift angle, DC bus voltage and inductor current of Level 1 charging.}
\end{figure}

\begin{figure}[!ht]
\centering
\includegraphics[width=1.1\textwidth]{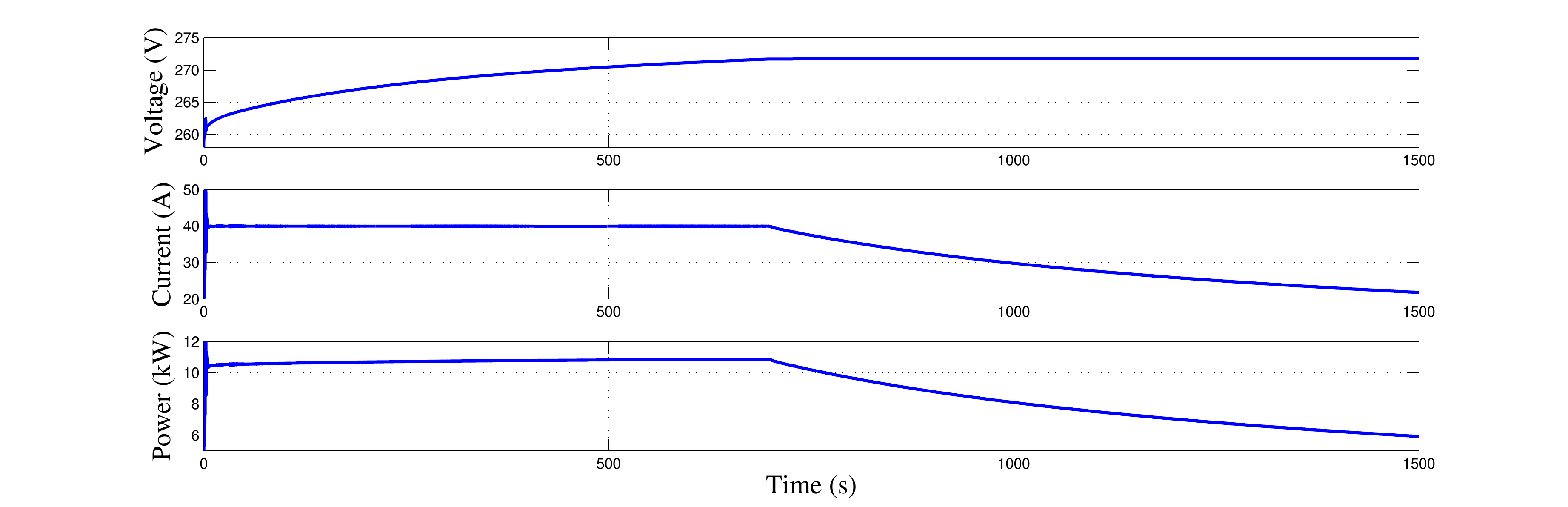}
\caption{\label{22}Current, voltage and charging power of Level 2 charging.}
\end{figure}

\begin{figure}[!ht]
\centering
\includegraphics[width=1.1\textwidth]{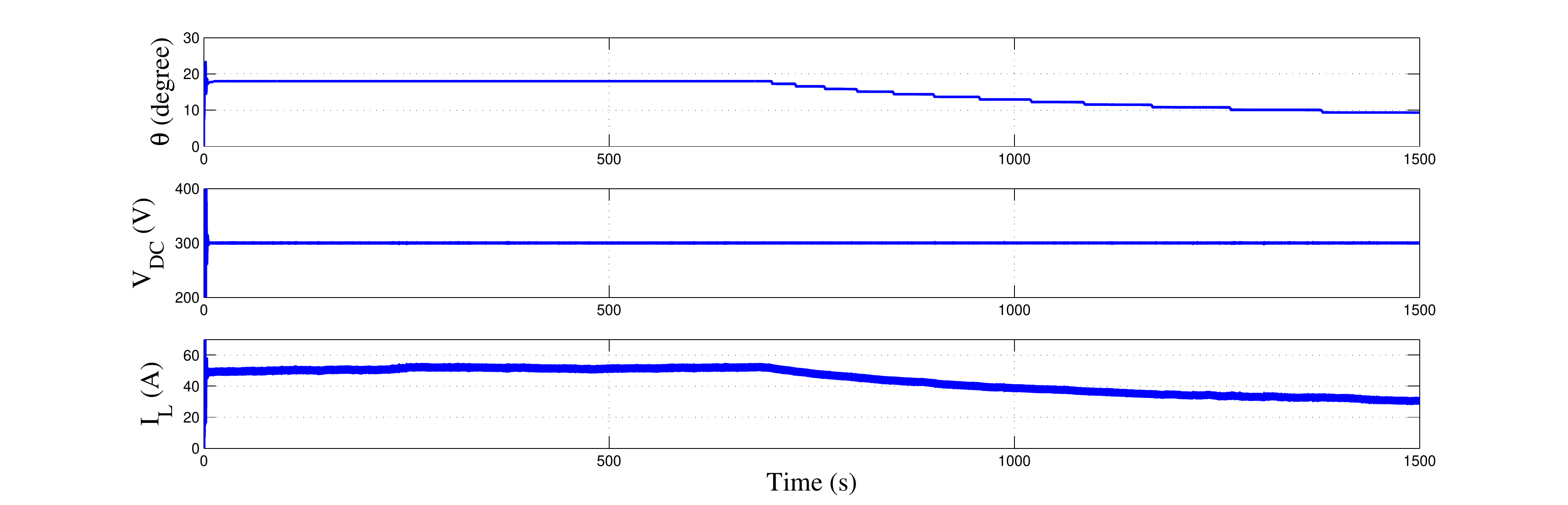}
\caption{\label{23}Phase shift angle, DC bus voltage and inductor current of Level 2 charging.}
\end{figure}

\begin{figure}[!ht]
\centering
\includegraphics[width=1.1\textwidth]{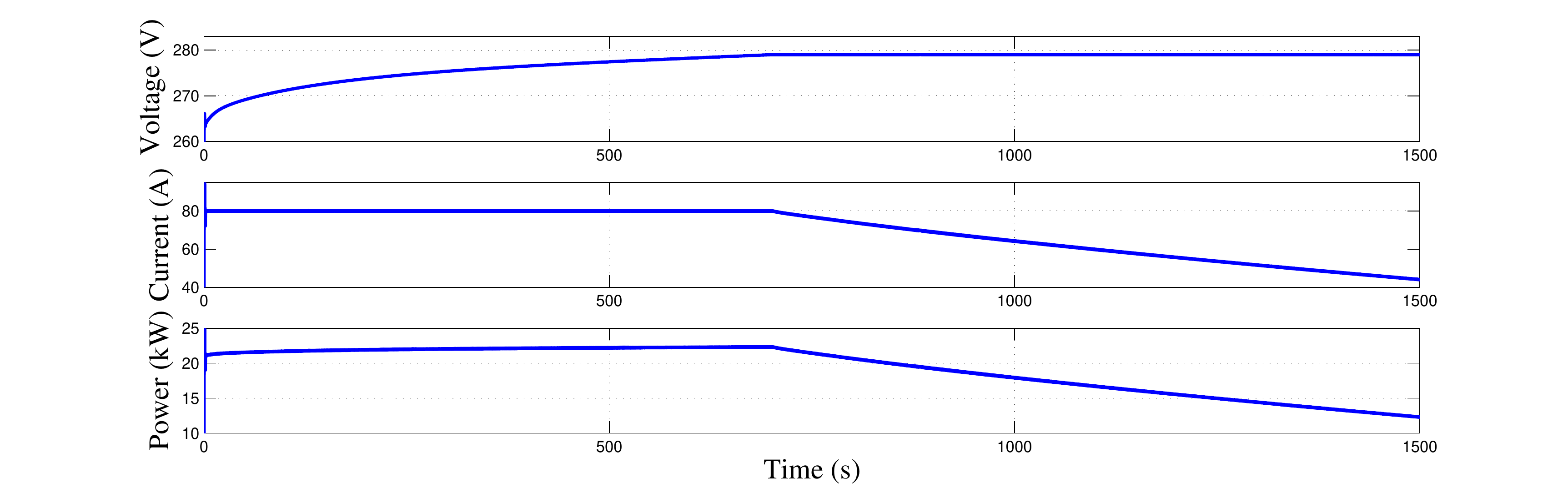}
\caption{\label{32}Current, voltage and charging power of Level 3 charging.}
\end{figure}

\begin{figure}[!ht]
\centering
\includegraphics[width=1.1\textwidth]{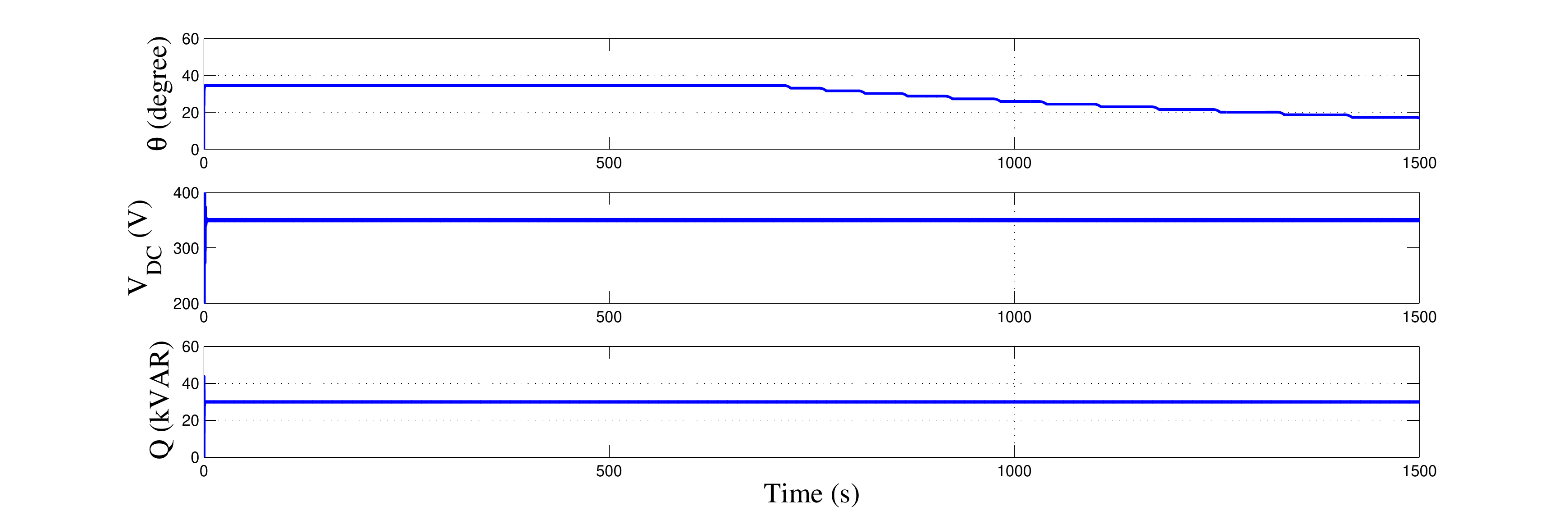}
\caption{\label{33}Phase shift angle, DC bus voltage and ractive power of Level 3 charging.}
\end{figure}

Figs. \ref{12}-\ref{13}  present Level 1 charging process. At the beginning, the charging starts
with a constant current, $5$ $A$, while the voltage is increasing. The current will decrease when
the voltage reaches the preset value, $262$ $V$. The DC bus for Level 1 charging is 300 $V$. The
results validate the CC/CV charging method. Figs. \ref{22}-\ref{23} show Level 2 charging
simulation, which also includes two stages: CC and CV. At CC stage, the current is 40 $A$ and
charging turns to CV mode when the voltage is $272$ $V$, and the DC bus voltage is also $300$ $V$. In CV mode, with the charging power decreasing, the phase shift and inductor are also followed to decrease. Figs. \ref{32}-\ref{33} present Level 3 charging process. The constant current is $80$ $A$ and the
constant voltage is $273$ $V$. The DC bus voltage is $350$ $V$ and the reactive power keeps as $30$
k$VAR$.

\begin{figure}[!ht]
\vspace{-0.15in}
\centering
\includegraphics[width=1.1\textwidth]{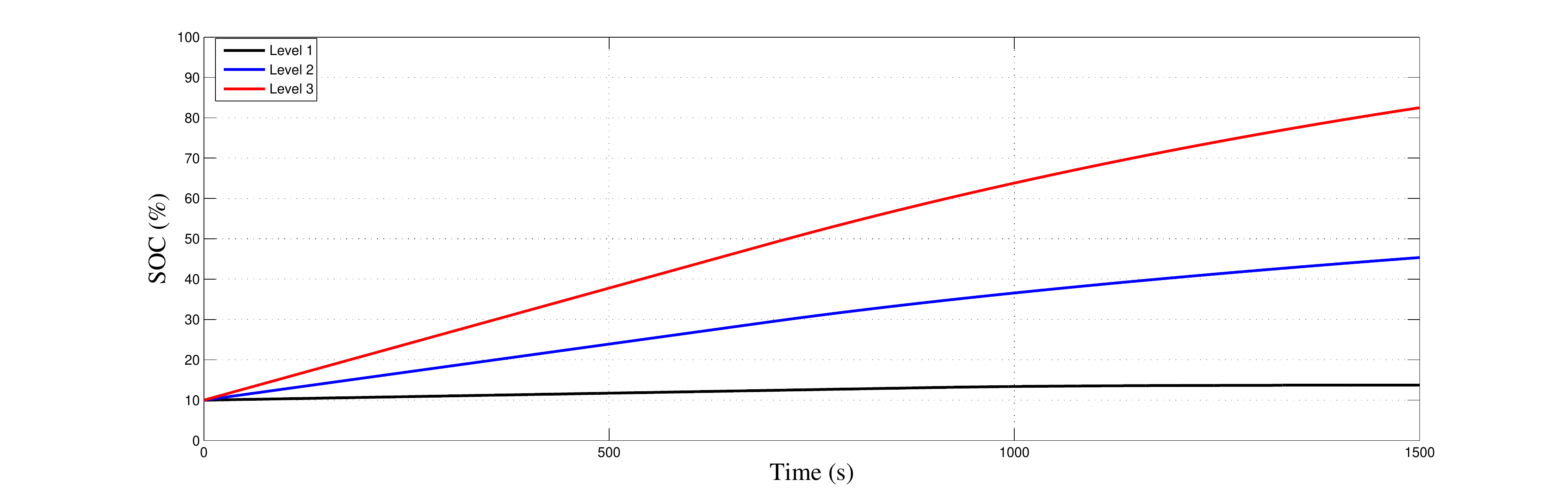}
\caption{\label{soc}SOC comparison of three testbeds.}
\vspace{-0.15in}
\end{figure}

Fig. \ref{soc} shows the SOC comparison of the three-level charging system in the same time period.
Level 3 charger is the fastest charging which increases the SOC from $10\%$ to $82\%$. Level 2
charging increases the SOC from $10\%$ to $48\%$. Level 1 charger is the slowest, and it only
charges the battery to be $5\%$ more in 25 minutes.

Based on the simulation waveforms, the numerical results are summarized in Table. \ref{results}.
The charging currents, voltages and power ratings correspond to the SAE's standard. The estimate
fully charging time means the time consuming to fully charge a $0\%$ SOC battery to $100\%$.

\begin{table}[h]
\caption{ Simulation analysis}
\label{results}
\begin{center}
\begin{tabular}{cccc}
   \hline\hline
     &Level 1 &Level 2 & Level 3\\
   \hline
   Initial SOC&$10\%$ & $10\%$ & $10\%$\\
    Initial phase shift & $12^o$ & $18^o$ & $33^o$\\
    DC bus voltage & $300$ V & $300$ V & $350$ V\\
    Charging power & $1.31$ kW & $10.92$ kW & $22.32$ kW\\
    Constant current& $5$ A & $40$ A & $80$ A\\
    Constant voltage& $262$ $V$ & $272$ $V$& $279$ $V$\\
    End SOC& $14.3\%$ & $48.4\%$& $82.3\%$\\
    \multirow{1}{*}{Estimated fully} &{10 hours} & {1.1 hour}&{35 mins}\\
    {charging time}& & &\\
    \hline\hline
\end{tabular}
\end{center}
\vspace{-0.15in}
\end{table}

\begin{figure}[!ht]
    \centering
    \subfigure[]
    {
        \includegraphics[width=0.5\textwidth]{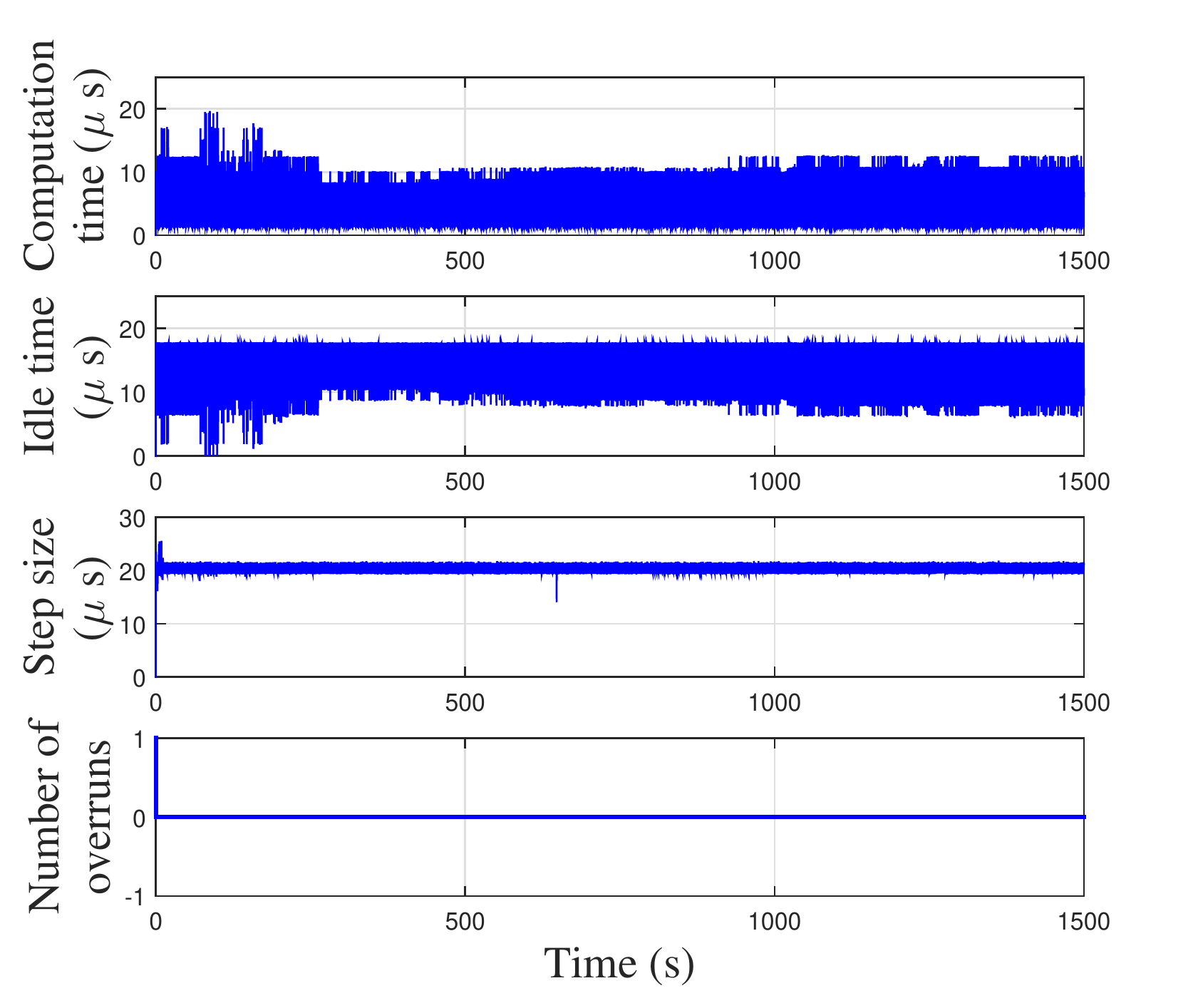}
    }
    \\
    \subfigure[]
    {
        \includegraphics[width=0.4\textwidth]{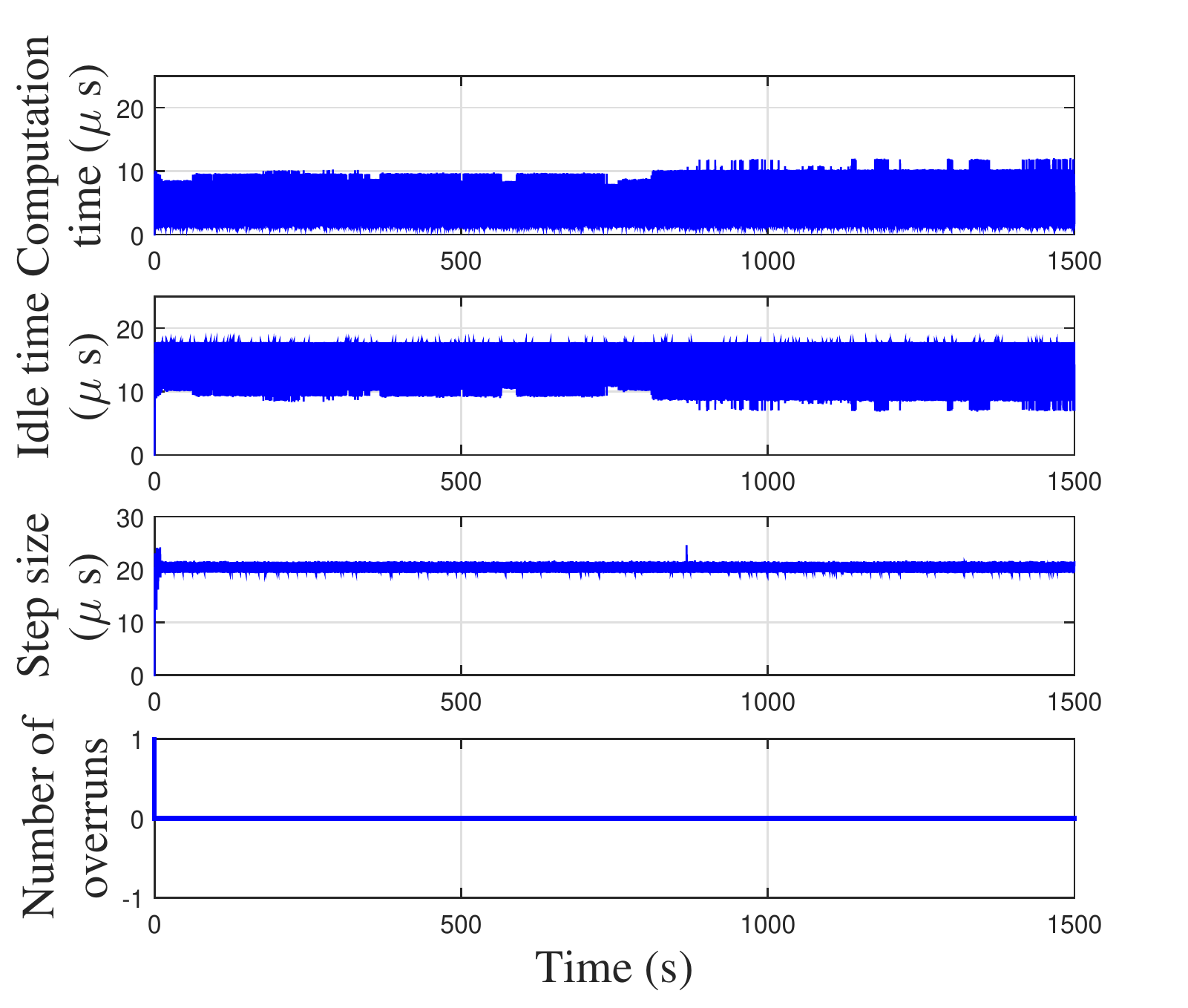}
    }
    \subfigure[]
    {
        \includegraphics[width=0.4\textwidth]{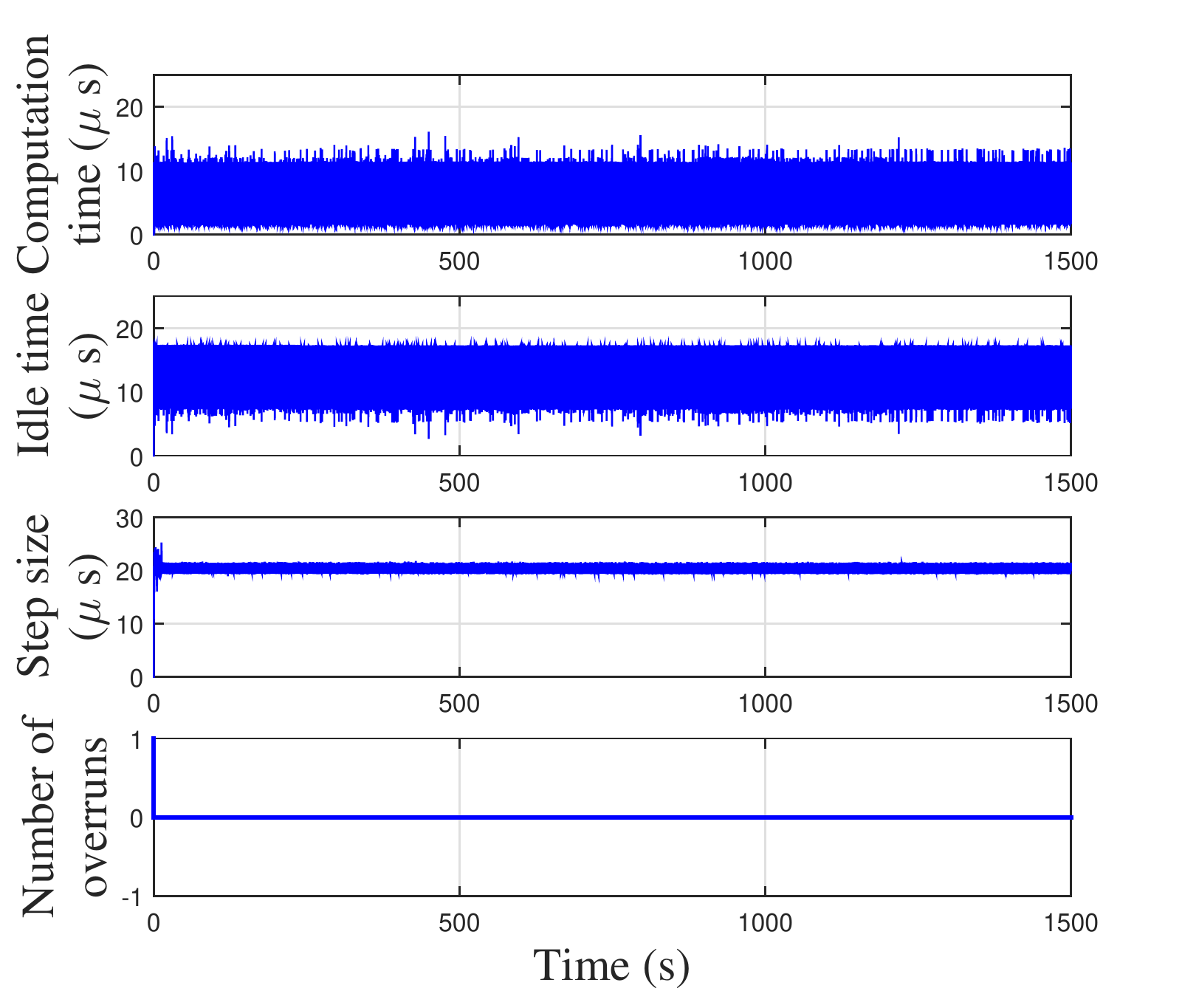}
    }
    \caption{RT-Lab performance for: (a) Level 1 charging, (b) Level 2 charging, (c) Level 3 charging.}
    \label{rtper}
    \vspace{-0.25in}
\end{figure}

In RT-Lab, Opmonitor block is used to record the model's procedures. Figs. \ref{rtper} show the
RT-Lab performance of three systems. Step sizes of all models are set as 20 $\mu s$, and no overrun
occurs in simulation period, which indicates stable simulation condition.

\section{Conclusion}
The paper presents the circuit topology, control and real-time simulation implementation of
charging systems for EVs at three levels. Level 1 and Level 2 battery charging systems consist of a
diode-based AC/DC converter, a PFC boost circuit, a DAB converter, a battery, and the related
control systems. A PFC controller is employed to ensure a constant DC bus voltage and unity power
factor. The CC/CV charging control is implemented the DAB converter. Level 3 charging system
consists of a bi-directional three-phase AC/DC voltage source converter and a DAB converter.
$V_{DC}/Q$ control is employed on the AC/DC converter to regulate the DC bus voltage and reactive
power. Simulations are implemented in RT-Lab simulator to demonstrate the charging processes
of the charging system performance. 





\bibliography{bio}




\end{document}